\documentclass[10pt,aps,prl,twocolumn,superscriptaddress,floatfix,secnumarabic]{revtex4-2}
\usepackage{amsmath,amssymb,amsfonts}
\usepackage{graphicx}
\usepackage{dcolumn}
\usepackage{bm}
\usepackage{xcolor}
\usepackage{physics}  
\usepackage{bbm}      
\usepackage{microtype}
\usepackage{colortbl}
\usepackage{booktabs}
\usepackage{url}
\usepackage{hyperref}
\hypersetup{
    colorlinks=true,
    citecolor=magenta,
    linkcolor=blue,    
    urlcolor=blue,
}

\usepackage{tikz} 
\usetikzlibrary{positioning,arrows.meta,calc}

\newcommand{\phdagger}{{\phantom{\dagger}}}

\newcommand{\papertitle}{Breaking the symmetry of collective enhancements in ensembles of emitters}

\begin{document}
\allowdisplaybreaks
\setlength{\textfloatsep}{10pt plus 2pt minus 2pt}
\setlength{\dbltextfloatsep}{10pt plus 2pt minus 2pt}
\setlength{\floatsep}{8pt plus 2pt minus 2pt}
\setlength{\intextsep}{8pt plus 2pt minus 2pt}
\title{\papertitle}

\author{Francesco Campaioli}
\email{francesco.campaioli@rmit.edu.au}
\affiliation{School of Science, RMIT University, Victoria, 3000, Australia}
\affiliation{RMIT Applied Quantum Technologies Centre, RMIT University, Melbourne, 3000, Victoria, Australia}

\author{Daniel E. G\'omez}
\email{daniel.gomez@rmit.edu.au}
\affiliation{School of Science, RMIT University, Victoria, 3000, Australia}

\date{\today}

\begin{abstract}
Collective interactions allow ensembles of emitters to exchange energy with light as a single coherent object, as in Dicke superradiance. However, collectivity enhances radiative loss and energy conversion alike, leaving their branching ratio unchanged. An efficiency gain requires a structure, such as a cavity, that suppresses the wasteful channel so that only the useful one inherits an enhancement. We derive the resulting conversion efficiency in closed form and apply it to published data on plasmonic photochemistry, showing that it supports the hypothesis of collective hot-electron transfer.
\end{abstract}

\maketitle

\noindent
Collective effects~\cite{Reitz_PQ2022a} are well known to offer a powerful handle on the rates of energy conversion and transfer. In 1954, Dicke showed  that $N$ emitters coupled to a common radiation field can synchronize and emit as a whole, in the celebrated phenomenon of \textit{superradiance}~\cite{Dicke1954Coherence,Barotov_AOM2023a}. 
Conversely, the same coherence enables \textit{superabsorption}~\cite{Higgins_NC2014a}, in which an ensemble takes up 
energy far faster than the sum of its parts~\cite{Quach_SA2022a, Hymas_LSA2026a, Yang_NP2021a, Joshi_PRA2022a, Boehme_NP2025a}. A renewed interest in collective effects in photochemistry and charge transport has recently identified how collective couplings can also enhance the transport of delocalised electronic states, via an extension of Marcus theory known as \textit{supertransfer}~\cite{Taylor2018, Kushwaha2026}, although its experimental verification has so far remained elusive.

However, collective enhancements do not discriminate between useful and wasteful channels, accelerating radiative emission just as effectively as absorption~\cite{Higgins_NC2014a}. A structure that suppresses radiation into free space, such as an optical cavity, can break this symmetry~\cite{Garcia-Vidal_S2021m, Quach_SA2022a}. Yet two central questions remain open. How does the yield of a collective conversion channel scale with the number of emitters? And how can an experiment reveal the signature of collective conversion?

Here we resolve both questions. First, we show that collectivity does not alter the competition between useful conversion and collective radiation, because both rates scale identically with the number $N$ of emitters. It can nevertheless improve the total yield of a particular energy-conversion channel available to a system by outpacing local non-radiative loss. A net gain in efficiency requires the symmetry between channels to be broken by a structure that suppresses radiation into free space, a role played here by an optical cavity, while leaving the useful pathway intact, so that only the latter inherits a collective enhancement. We formulate an exact single-excitation theory that captures this mechanism and derive closed-form yields, together with their scaling with $N$ and the cooperativity threshold above which the cavity improves conversion. 

We then apply our theory to published data on plasmonic photochemistry and show that structural control breaks the symmetry between competing channels. The scaling of the yield with particle density supports collective hot-electron transfer and provides, to our knowledge, the first evidence in support of supertransfer in plasmonic energy conversion, through a simple test that requires only yields measured at different densities.

\begin{figure*}[t]
	\centering
	\includegraphics[width=\textwidth]{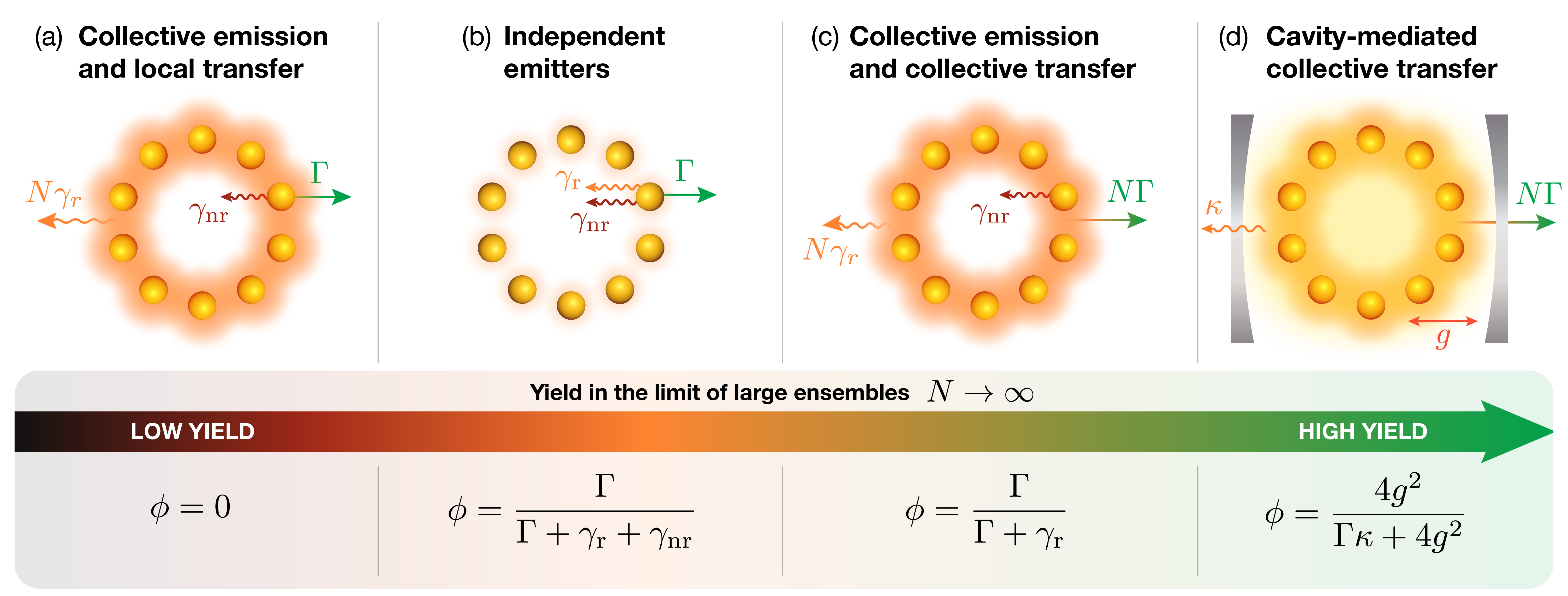}
\caption{\textbf{Breaking the symmetry between collective channels.} Illustration of ensembles of $N$ emitters (spheres) in which an excitation can be converted into charge at rate $\Gamma$ per emitter (green arrows), emitted as light at rate $\gamma_{\rm r}$ (orange wavy arrows) or lost as heat at rate $\gamma_{\rm nr}$ (dark red wavy arrows). (a)~With collective emission and local transfer, radiation is enhanced to $N\gamma_{\rm r}$ while conversion is not, and the yield vanishes. (c)~When transfer is also collective, conversion is enhanced to $N\Gamma$ by the same factor as radiation. Both outpace the heat loss, but their ratio is fixed, so the yield saturates at $\Gamma/(\Gamma+\gamma_{\rm r})$. (d)~A cavity with loss rate $\kappa$, coupled to each emitter with strength $g$, replaces collective radiation by coherent exchange with the cavity mode. Light is then lost only at the fixed rate $\kappa$, while conversion retains its collective enhancement.}
	\label{fig:mechanism}
\end{figure*}

\paragraph*{Collective emitters and conversion channels.} We consider a minimal model of an ensemble of $N$ identical emitters, each coupled identically to a single cavity mode in a star topology with the cavity as the hub.
We describe the dynamics of this system using the Tavis--Cummings Hamiltonian in the rotating-wave approximation~\cite{Tavis_PR1968a}, with $\hbar\equiv1$:
\begin{equation}
	H = \omega_c \hat{a}^\dagger \hat{a} +\omega_0\sum_{m=1}^{N}\hat{p}_m^\dagger \hat{p}_m +g\sum_{m=1}^{N} \left(\hat{a}^\dagger \hat{p}_m+\hat{a} \hat{p}_m^\dagger\right),
	\label{eq:H}
\end{equation}
where $\hat{a}$ ($\hat{a}^\dagger$) annihilates (creates) a cavity photon of frequency $\omega_c$, and $\hat{p}_m$ ($\hat{p}_m^\dagger$) annihilates (creates) an excitation on emitter $m$ of frequency $\omega_0$. The single-emitter cavity coupling $g$ is set by the cavity geometry and, for a fixed cavity, is independent of $N$. 

We then make a further assumption, by considering low excitation density in the emitters, meaning that the number of excitations in the array at any time is much smaller than $N$. Under this assumption, dynamics remain approximately confined to the single-excitation manifold. The symmetric structure of the coupling means that only one collective linear combination of the emitters, known as the \textit{bright mode} $\hat{b} = N^{-1/2}\sum_{m=1}^{N}\hat{p}_m$, couples to the cavity with effective strength $G = g\sqrt{N}$. We denote by $|{\rm vac}\rangle$ the ground state, with no photon in the cavity and no excitation in the emitters, by $|1_c\rangle = \hat{a}^\dagger|{\rm vac}\rangle$ the state with a single cavity photon, and by $|B\rangle = \hat{b}^\dagger|{\rm vac}\rangle = \sum_{m}\hat{p}_m^\dagger|{\rm vac}\rangle/\sqrt{N}$ the state with a single excitation in the bright mode. The remaining $N-1$ orthogonal superpositions, known as \textit{dark states}, decouple entirely and play no role in the absence of dephasing and disorder.

We then model irreversible loss and energy conversion using
Lindblad operators $L_k$ representing each channel~\cite{Carmichael_2013a, Breuer2002, Campaioli2024a}, for all cases illustrated in Fig.~\ref{fig:mechanism}~(a)--(d).
The wasteful loss channels are local non-radiative relaxation ($\sqrt{\gamma_{\rm nr}}\hat{p}_m$ per emitter $m$) for all the considered cases, radiative decay, which is local ($\sqrt{\gamma_{\rm r}}\hat{p}_m$ per emitter $m$) for independent emitters and collective ($\sqrt{\gamma_{\rm r}}\sum_{m}\hat{p}_m = \sqrt{N\gamma_{\rm r}}\hat{b}$) for emitters sharing the free-space radiation field, and cavity loss ($\sqrt{\kappa}\hat{a}$), which replaces free-space radiation when the emitters are coupled to the cavity. The useful energy conversion channel can be either a local capture ($\sqrt{\Gamma}\hat{p}_m$ per emitter) or collective capture ($\sqrt{\Gamma}\sum_{m}\hat{p}_m = \sqrt{N\Gamma}\hat{b}$), acting coherently on the bright mode. 

The observable of interest is the probability that an excitation eventually leaves the system through the useful conversion channel,
\begin{equation}
	\phi = \int_0^\infty dt\,
	\langle\widetilde{\psi}(t)| L_\Gamma^\dagger L^\phdagger_\Gamma
	|\widetilde{\psi}(t)\rangle .
	\label{eq:yield}
\end{equation}
Here $|\widetilde{\psi}(t)\rangle$ is the non-normalised no-jump state, obtained by evolving the initial state $|\psi_0\rangle$ under the effective non-Hermitian Hamiltonian
$H_{\rm eff} = H - \frac{i}{2}\sum_k L_k^\dagger L_k^\phdagger$~\cite{Plenio_RMP1998a}.
Because every channel removes the excitation, this description is exact in the single-excitation sector. All yields can be evaluated using the Lyapunov equation~\cite{Purkayastha_PRA2022a}. Here, we define yields per photon absorbed by a considered structure, unless specified otherwise. Details are given in Appendix~\ref{sm:lyapunov-framework}.

\paragraph*{Independent emitters.} We begin by establishing a baseline for independent emitters. Each emitter converts energy at rate $\Gamma$, radiates at rate $\gamma_{\rm r}$ and loses energy non-radiatively at rate $\gamma_{\rm nr}$, through channels that act on that emitter alone. Because these local channels respond only to whether an emitter is excited, and not to the phase relations between emitters, every relaxation event is a conversion with probability $\phi_{\rm ind} = \Gamma/(\Gamma+\gamma_{\rm r}+\gamma_{\rm nr})$. 

\begin{figure}
	\centering
	\includegraphics[width=0.9\columnwidth]{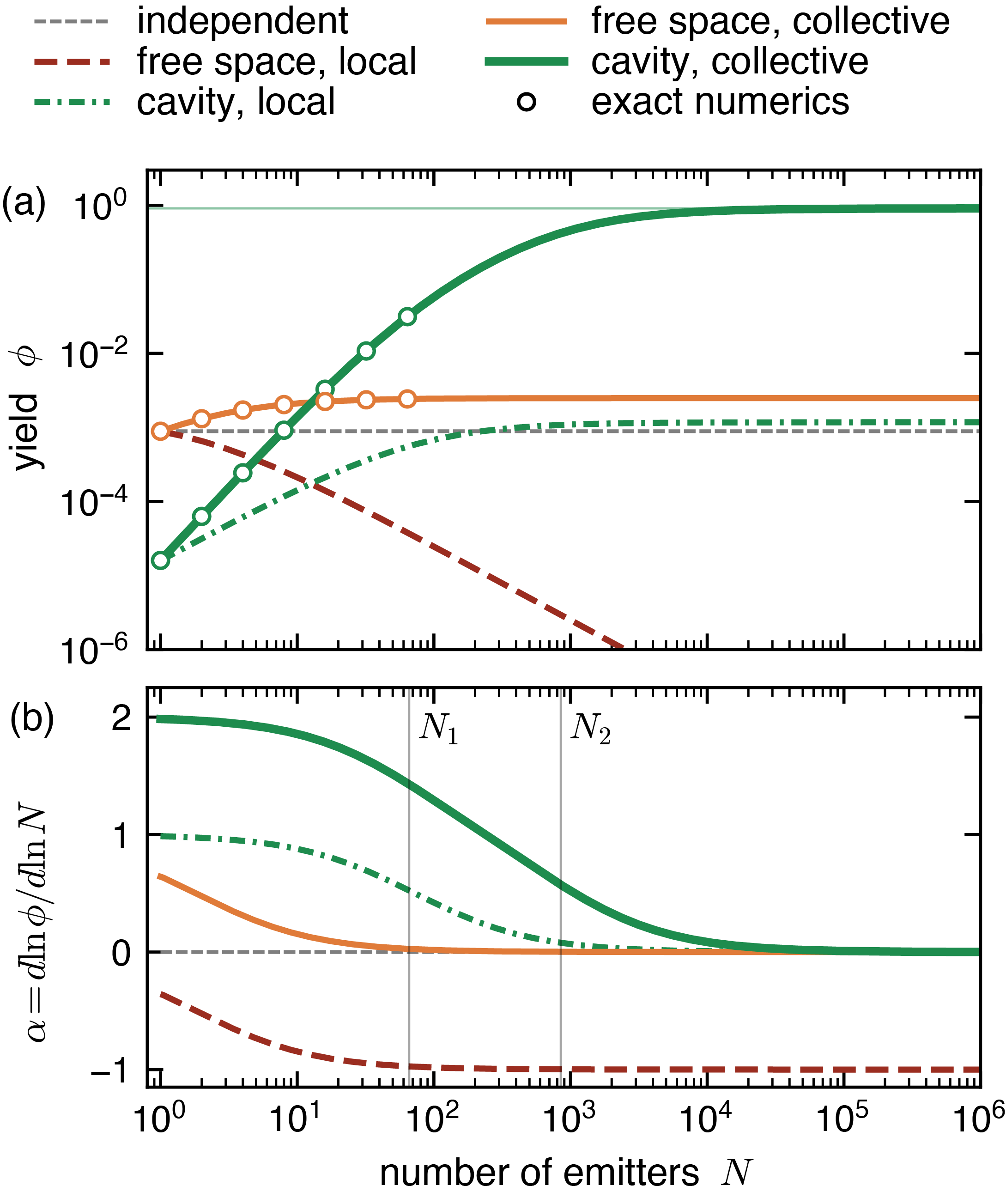}
	\caption{\textbf{Scaling of energy conversion yield.} (a)~Energy conversion yield $\phi$ as a function of the number $N$ of emitters, for independent emitters (gray dashed line), emitters in free space with local (dark red dashed line) and collective (orange line) conversion, and emitters coupled to a cavity with local (green dashed-dotted line) and collective (green line) conversion. Open circles are exact numerical solutions of the full $(N+1)$-dimensional problem for $N\le64$, which coincide with the closed-form yields. (b)~Local scaling exponent $\alpha=d\ln\phi/d\ln N$ for the same cases. Vertical lines mark the scales $N_1$ and $N_2$ at which the absorption and conversion factors of Eq.~\eqref{eq:phi_cavity} saturate. For all cases 
    $\Gamma=0.61$~THz, $\gamma_{\rm r}=243$~THz, $\gamma_{\rm nr}=441$~THz, $\kappa=76$~THz and large-ensemble cooperativity $C_\infty=4g^2/(\kappa\Gamma)=10$.}
	\label{fig:yield}
\end{figure}

\paragraph*{Symmetric collective enhancements.} We now consider emitters that absorb and emit via a delocalised bright state. The collective bright mode decays via useful conversion at rate $\Gamma$ (local) or $N\Gamma$ (collective), radiative loss at rate $N 
\gamma_{\rm r}$, and non-radiative loss at rate $\gamma_{\rm nr}$. When the useful conversion channel is \emph{local} 
the energy conversion yield $\phi_{\rm fs}^{\rm loc} = \Gamma/(\Gamma+N\gamma_{\rm r}+\gamma_{\rm nr})$ which scales as $N^{-1}$, vanishing in the limit of large ensemble size, as shown in Fig.~\ref{fig:yield}~(a). When the useful conversion channel acts \emph{collectively}, i.e., through a shared mode, both conversion and radiation scale with the size of the ensemble $N$, as shown in Fig.~\ref{fig:mechanism}~(c). Because they are enhanced identically, they compete symmetrically. In this case the yield reads $\phi_{\rm fs}^{\rm col} =N\Gamma/[N(\Gamma +\gamma_{\rm r})+\gamma_{\rm nr}]$ and, at large $N$, it saturates at 
$\phi_{\rm fs}^{\rm col} = \Gamma/(\Gamma + \gamma_{\rm r})$.

\paragraph*{Breaking the symmetry.} Favouring conversion over loss channels requires an external structure that removes the collective enhancement of the lossy channel, while leaving the useful one intact. The cavity provides exactly this asymmetry, as illustrated in Fig.~\ref{fig:mechanism}~(d). Instead of radiating into free space, the bright mode exchanges energy coherently with the cavity mode, so light emitted by the array remains in the cavity and can be reabsorbed. Energy is then lost as light only when it escapes the cavity, at a rate $\kappa$ that does not depend on $N$. The radiative loss therefore no longer grows with the size of the ensemble, whereas the conversion channel, acting on the same bright mode, retains its collective enhancement $N\Gamma$.

For cavity-mediated energy conversion, the problem reduces to two coupled degrees of freedom, the cavity and the bright mode, which exchange energy at the collective rate $G=g\sqrt{N}$. The cavity loses energy at rate $\kappa$, and the bright mode at rate $\gamma_{\rm nr}+N\Gamma$. The energy conversion yield reflects the competition between these two effective degrees of freedom, factorising as
\begin{equation}
	\phi_{\rm cav}^{\rm col}(N) = \frac{4G^2}{\kappa(\gamma_{\rm nr}+N\Gamma)+4G^2}
	\cdot\frac{N\Gamma}{\kappa+\gamma_{\rm nr}+N\Gamma}.
	\label{eq:phi_cavity}
\end{equation}
The first factor equals $C_N/(1+C_N)$, where $C_N = 4G^2/[\kappa(\gamma_{\rm nr}+N\Gamma)]$ is the collective cooperativity, and measures how effectively the ensemble draws the photon out of the cavity before it leaks. The second factor is the share of conversion among all the losses of the coupled system.

The scaling exponent $\alpha(N) = d\ln\phi/d\ln N$ decreases continuously from 2 to 0, as shown by the green line in Fig.~\ref{fig:yield}~(b). At small $N$ both factors are small and each grows linearly with $N$, so $\alpha \approx 2$. The exponent passes through $\alpha \approx 3/2$ near $C_N \approx 1$, where the array begins to absorb the photon as fast as the cavity loses it. 

\paragraph*{Saturation and impedance matching.} 
The yield of Eq.~\eqref{eq:phi_cavity} has no optimum in $N$. Near $N_2 = (\kappa+\gamma_{\rm nr})/\Gamma$ collective conversion overtakes cavity escape and non-radiative decay, the conversion factor approaches one, and the exponent falls towards zero. Beyond this point the bright mode is drained faster than the cavity can feed it, the polaritons disappear, and adding emitters no longer changes the yield
\begin{equation}
	\lim_{N\to\infty}\phi_{\rm cav}^{\rm col}(N) = \frac{4g^2}{4g^2+\kappa\Gamma},
	\label{eq:phi_asymp}
\end{equation}
which depends on microscopic coupling $g$ and the cavity loss $\kappa$. This yield is set by the competition between excitation transfer from the cavity to the ensemble of emitters, at rate  $4g^2/\Gamma$, and escape from the cavity, at rate $\kappa$; see Appendix~\ref{sec:methods_cavity_yield} for details.

Let us note that the asymptotic yield exceeds that of independent emitters if and only if 
\begin{equation}
	\frac{4g^2}{\kappa\Gamma} > \frac{\Gamma}{\gamma_{\rm r}+\gamma_{\rm nr}}.
	\label{eq:coop_threshold}
\end{equation}
In terms of the single-emitter cooperativity $C = 4g^2/[\kappa(\gamma_{\rm r}+\gamma_{\rm nr})]$, the condition reads $C > [\Gamma/(\gamma_{\rm r}+\gamma_{\rm nr})]^2$. For comparison, cavity-coupled emitters with a local conversion channel benefit only from the cooperativity factor in Eq.~\eqref{eq:phi_cavity}, while the conversion factor becomes $\Gamma/(\kappa+\gamma_{\rm nr}+\Gamma)$ and no longer grows with $N$, as shown in Fig.~\ref{fig:yield}~(a). In these settings, superlinear growth of the single-photon conversion yield, requires three conditions simultaneously: a delocalised excitation, a collective conversion channel, and a cavity that removes the collective radiative loss. Removing any one of them limits the growth of the yield to at most linear in $N$. Therefore, at fixed $g$, observing $\alpha > 1$ in the weak-pump limit is an experimental test that conversion is collective. 

\begin{figure}
	\centering
	\includegraphics[width=\columnwidth]{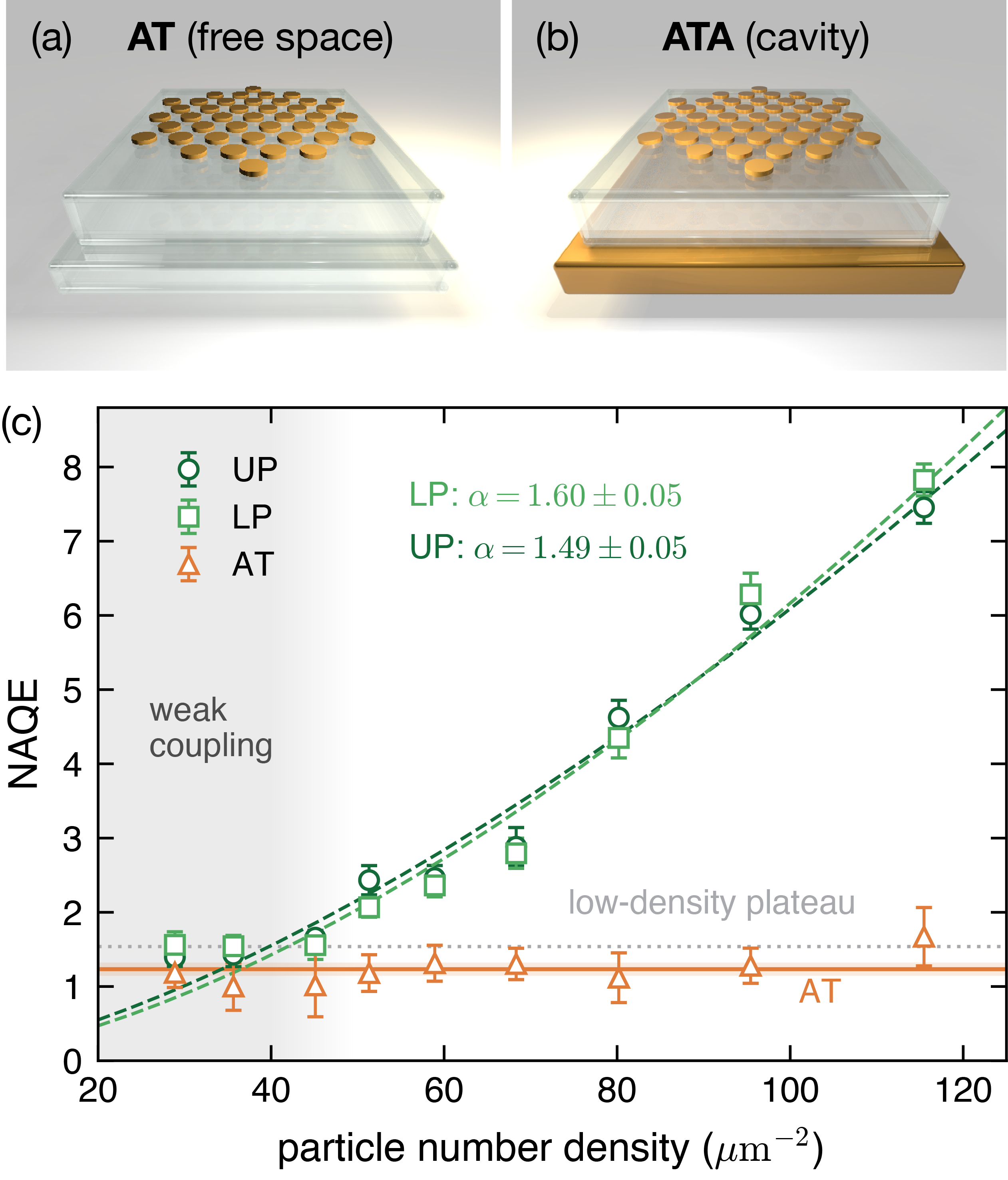}
	\caption{\textbf{Measured conversion efficiency with and without the cavity.} Schematic of the (a)~free space (AT) and (b)~cavity-coupled (ATA) geometries. (c)~Normalised apparent quantum efficiency (NAQE) as a function of particle number density (PND) for the upper polariton (UP, dark green circles), lower polariton (LP, light green squares) and AT (orange triangles) samples. The gray dotted line is the low-density plateau, a single level fitted to both polariton branches below PND~$=50\,\mu{\rm m}^{-2}$. The curves are power-law fits to the UP and LP branches (dashed lines), with ${\rm NAQE}\propto{\rm PND}^{\alpha}$, extended over the full range. 
    }
	\label{fig:AQEvsN}
\end{figure}

\paragraph*{Experimental realisation with plasmonic photochemistry.}
To test the general mechanism proposed above, we study the plasmonic platform previously realised experimentally by~\citet{Liu_AN2023a}, who kindly provided us with their data. We quote energies and couplings in eV and decay rates in THz.
The structures consist of periodic hexagonal arrays of Au nanodisks deposited on TiO$_2$, with the particle number density (PND) varied while the nanoparticle dimensions remain fixed (Fig.~\ref{fig:AQEvsN}). For these nanoparticles, $\gamma_{\rm r} \approx 240$~THz and $\gamma_{\rm nr} \approx 440$~THz~\cite{Forcherio_AOM2017a}. The number of particles in the illuminated mode is $N = A\cdot{\rm PND}$, where $A$ is the mode area.

The experimental data of Ref.~\cite{Liu_AN2023a} provide transient absorption for two structurally different geometries. In the cavity-free geometry (AT) the nanoparticle array was deposited directly on TiO$_2$ supported by an optically-transparent substrate, shown in Fig.~\ref{fig:AQEvsN}~(a).

In the cavity geometry (ATA), an additional Au mirror beneath the TiO$_2$ spacer forms a Fabry--P\'erot cavity, shown in Fig.~\ref{fig:AQEvsN} (b). The cavity hybridises with the collective plasmon to form polaritonic modes~\cite{Shi_NN2018a}, as shown in Fig.~\ref{fig:steadystatespectra}, while simultaneously replacing free-space superradiant emission by cavity leakage $\kappa$.

The measurement protocol is summarised in Appendix~\ref{app:em}. Because AT and ATA structures exhibit different optical absorptions, the transient signals are normalised by the number of absorbed photons to define an apparent quantum efficiency (AQE)
\begin{equation}
\label{eq:apparent_quantum_yield}
    \mathrm{AQE} = \frac{\Delta\mathrm{OD}_{\max}}{n \cdot \mathrm{Abs}},
\end{equation}
where $n$ is the incident photon count and $\mathrm{Abs}$ is the measured absorption at the pump wavelength. Although not an absolute quantum yield, the AQE is proportional to the probability that a photon taken up by the structure generates an injected electron. Fig.~\ref{fig:AQEvsN}~(c) shows the measured AQE for both AT and ATA samples as a function of particle number density, where all the data were supplied to us normalised to the AT case for PND$\sim 35.6 \mu{\rm m}^{-2}$. We denote this normalised quantity by ${\rm NAQE}:={\rm AQE}/{\rm AQE}_{\rm AT}{\rm(35.6 \mu m^{-2})}$, which has the same density dependence as AQE.

In the AT samples the absorbance is measured in transmission, so it includes photons that the array re-radiates. The AQE therefore counts re-radiation as a loss, corresponding to the free-space collective yield $\phi_{\rm fs}^{\rm col}$. In the mirror-backed ATA geometry, the measured absorbance is $1-R$, where $R$ is the reflectance, and light re-emitted through the input side leaves as reflection and is not counted. We therefore model the experiment as a steady drive at the frequency of the pumped polariton, with the yield defined per photon dissipated in the structure. The coupling of the cavity to the incoming light then drops out, and the only competitor of the array for a photon in the cavity is parasitic absorption in the mirror, at rate $\kappa_0$. The resulting yield is
\begin{equation}
	\phi_{\rm ATA} = \frac{N_{\rm coh}\Gamma\,G^2}{\kappa_0\left(\delta^2+\frac{\gamma_B^2}{4}\right)+\gamma_B G^2},
	\label{eq:phi_drive}
\end{equation}
where $\gamma_B=\gamma_{\rm nr}+N_{\rm coh}\Gamma$, $\delta = \omega - \Omega_B$ is the detuning of the pump frequency $\omega$ from the bright mode; see Appendix~\ref{sm:driven} for details. 

Since the two polaritons differ only in their detuning $\delta$, the equality of their yields bounds 
$\kappa_0<4.4$~THz, about $2\%$ of $\kappa$ so that $\phi_{\rm ATA}\approx N_{\rm coh}\Gamma/\gamma_B$. Here, $1\leq N_{\rm coh}\leq N$ is the number of particles over which the bright mode transfers electrons coherently, representing the extent of delocalisation of single excitation over the emitters.

\paragraph*{Predictions and comparison with experiment.} 
The NAQE of the AT structure of Fig.~\ref{fig:AQEvsN}~(a)  remains essentially constant across the full particle-density range, with a mean of $1.23\pm 0.09$ and a fitted exponent of $\alpha = 0.16\pm 0.17$, in general consistent with the scaling predicted by independent emitters and symmetric collective enhancements.
However, increasing PND redshifts the plasmon resonance and increases its measured linewidth. A nearest-neighbour coupling model reproduces the redshift; see Appendix~\ref{sec:experiment_SI} for details. This indicates a delocalised bright mode on the hexagonal array, and with it a physical footing for collectively enhanced radiative loss, $N_{\rm coh}\gamma_r$, even without a cavity. 
If hot-electron injection is local to each particle, the AQE would \emph{decrease} with $N$ as collective radiative loss increasingly outcompetes an unenhanced conversion channel. Instead, if injection is itself collective, the AQE should trend to a flat asymptote with increasing $N$. The data show the latter, offering the experimental signature that both the useful and wasteful channels scale equally with $N$.

By contrast, the cavity ATA structures outperform AT in NAQE, an advantage that grows sharply with increasing density. A power law fitted over all densities, ${\rm NAQE}\propto{\rm PND}^\alpha$, gives $\alpha = 1.49\pm 0.05$ for the upper polariton (UP) and $\alpha = 1.60\pm 0.05$ for the lower polariton (LP), as shown in Fig.~\ref{fig:AQEvsN}~(c). However, to reject the local electron capture hypothesis one must proceed carefully, since the superlinear growth $G^2\propto{\rm PND}^{1.46}$, shown in Fig.~\ref{fig:G_hypothesis}~(a), can in principle support a growing yield even for local transfer.
To do so, we separate the transfer rate from the optical factors in the yield, as detailed in Appendix~\ref{app:em}. Local transfer then predicts a rate that is constant in density, whereas collective transfer predicts one that grows. The data reject local transfer, with $p\approx4\times10^{-25}$ (UP) and $8\times10^{-22}$ (LP). Collective transfer rate grows as ${\rm PND}^{1.7}$ for both polaritons, giving $p\approx4\times10^{-3}$ and $9\times10^{-4}$, an improvement of about twenty orders of magnitude.

\paragraph*{Discussion.} The mechanism behind the collective enhancement of electron transfer is not obvious \textit{a priori}. Conventional plasmonic understanding invokes local interfacial photoinduced charge transfer~\cite{Brongersma_NN2015a,Hartland_AEL2017a,Cortes_AN2020a}, and each electron is indeed injected from a single nanoparticle. What is collective is the excitation that feeds the injection. In the cavity structures the plasmon is delocalised over many nanoparticles, as proposed by Liu \textit{et al.}~\cite{Liu_AN2023a}, so that it can reach the particles with the fastest injection, which a localised excitation cannot do. A purely local mechanism, acting on an excitation confined to one particle, could depend on particle number only through the cooperative factor, as shown in Table~\ref{tab:yields}. Nor can the rise be attributed to near-field electromagnetic hot-spot enhancement activated as particles are brought closer together~\cite{Boerigter_AN2016a,Foerster_AN2017a}. Any such effect would be present equally in the AT structures, which share the same array and density series, yet show no corresponding rise. The observed increase of AQE with particle number is therefore consistent with injection driven by a delocalised plasmonic excitation.

The predicted enhancement depends critically on maintaining the phase coherence defining the bright collective mode. Local dephasing at rate $\gamma_\phi$ transfers population into dark states and suppresses the collective enhancement unless $\gamma_\phi\ll\gamma_B$. In the plasmonic realisation, the polariton splitting places the lower polariton well below the band of dark plasmon modes, by more than its linewidth, while the upper polariton lies just above this band; see Appendix~\ref{sm:dephasing}.

\paragraph*{Conclusions.}
Our analysis of the experiments of Ref.~\cite{Liu_AN2023a} supports the hypothesis that electron capture is enhanced by \textit{supertransfer}. In supertransfer, spatial delocalisation enhances donor--acceptor coupling through constructive interference of transfer amplitudes~\cite{Taylor2018,Kushwaha2025}. This is, to the best of our knowledge, the first evidence in support of supertransfer in plasmonic energy conversion. This work also establishes two further points. First, collective enhancement benefits energy conversion only when the symmetry between competing decay pathways is broken. Second, the closed-form yield provides a direct and simple experimental test of supertransfer, which here excludes local injection and is consistent with a coherent fraction of the array that grows with density.

The underlying mechanism is not restricted to plasmonic photochemistry. Any system containing (i) a collective bright mode, (ii) a useful decay channel sharing that collective symmetry, and (iii) an external structure capable of suppressing a competing loss channel can exhibit the same behaviour. Candidate platforms include cavity quantum electrodynamics systems~\cite{Walther_RPP2006a}, molecular aggregates~\cite{Barotov_AOM2023a}, superconducting circuits~\cite{Blais_RMP2021a} and quantum batteries~\cite{Campaioli_RMP2024a}, where an analogous competition between collective charging and collective self-discharge limits the achievable charging power. 

\begin{acknowledgments}
\paragraph*{Acknowledgements.} D.G. acknowledges support from the Australian Research Council (ARC) through Discovery Project DP230101764. The Authors acknowledge the use of Claude (Anthropic, Claude Opus 5.5, Fable 5.2, and previous versions) for language editing and for assistance in cross-checking analytical derivations and numerical results. All results were independently verified by the authors, who take full responsibility for the content of this manuscript. F.C. acknowledge access to Claude provided by Anthropic through the Claude Team Plan for Scientists.
\end{acknowledgments}

\paragraph*{Data availability.}
The data that support the findings of this article are not publicly available. The experimental data were provided by the authors of Ref.~\cite{Liu_AN2023a}. The data and the code used for the analysis are available from the authors upon reasonable request.

\bibliographystyle{apsrev4-2}
\bibliography{main}

\appendix
\setcounter{secnumdepth}{3}
\renewcommand{\thesubsection}{\thesection.\arabic{subsection}}
\section*{Appendix}

\noindent
Following the notation introduced in the main text, the cavity mode has frequency $\omega_c$, loss rate $\kappa$, and annihilation operator $\hat{a}$. Emitters $m=1,\dots,N$ have frequency $\omega_0$ and annihilation operator $\hat{p}_m$, couple to the cavity with strength $g$, convert energy at rate $\Gamma$, radiate at rate $\gamma_{\rm r}$ and relax non-radiatively at rate $\gamma_{\rm nr}$. The states $|{\rm vac}\rangle$, $|1_c\rangle$ and $|B\rangle$ are the vacuum state (no excitation), the one cavity photon state, and one excitation in the bright mode state, respectively. We set $\hbar=1$ and quote energies and couplings in eV and rates in THz.

\section{Lyapunov framework for single-excitation systems}
\label{sm:lyapunov-framework}
\noindent
All yields of the main text follow from a single algebraic equation, without time propagation~\cite{CaoSilbey2009,Gajic1995,Purkayastha_PRA2022a}. Each channel $k$ removes the excitation through a jump operator $L_k$ and leaves the system in $|{\rm vac}\rangle$. The single excitation is therefore described exactly by the non-normalised no-jump state $|\widetilde{\psi}(t)\rangle$ of Eq.~\eqref{eq:yield}, or by $\rho(t)=|\widetilde{\psi}(t)\rangle\langle\widetilde{\psi}(t)|$, which evolves as
\begin{equation}
	\dot\rho=-i\left(H_{\rm eff}\rho-\rho H_{\rm eff}^\dagger\right),
	\label{eq:nojump}
\end{equation}
generated by the effective non-Hermitian Hamiltonian 
\begin{equation}
    H_{\rm eff}=H-\tfrac{i}{2}\sum_k L_k^\dagger L_k.
    \label{eq:Heff_def}
\end{equation}
The probability that the excitation leaves through channel $k$ is the time-integrated flux into that channel,
\begin{equation}
	\phi_k=\int_0^\infty{\rm Tr}\left[L_k^\dagger L_k\,\rho(t)\right]dt={\rm Tr}\left(L_k^\dagger L_k M\right),
	\label{eq:phik}
\end{equation}
where 
\begin{equation}
M=\int_0^\infty\rho(t)\,dt.
\end{equation}
Integrating Eq.~\eqref{eq:nojump} over $[0,\infty)$ and using $\rho(\infty)=0$,
\begin{align}
	&\int_0^\infty\dot\rho\,dt=\rho(\infty)-\rho(0)=-\rho(0),\\
	&\int_0^\infty\dot\rho\,dt=-i\left(H_{\rm eff}M-MH_{\rm eff}^\dagger\right),
\end{align}
which together give the Lyapunov equation
\begin{equation}
	\rho(0)=iH_{\rm eff}M-iMH_{\rm eff}^\dagger,
	\label{eq:lyapunov}
\end{equation}
where $\rho(0)=|\psi_0\rangle\langle\psi_0|$. This equation is linear in $M$. All yields thus follow from the solution of one linear equation, satisfying the conservation rule $\sum_k \phi_k = 1$. Indeed, Eq.~\eqref{eq:Heff_def} gives $\sum_k L_k^\dagger L_k=i(H_{\rm eff}-H_{\rm eff}^\dagger)$, so that, by the cyclic property of the trace and Eq.~\eqref{eq:lyapunov},
\begin{align}
	\sum_k\phi_k&={\rm Tr}\left(iH_{\rm eff}M-iMH_{\rm eff}^\dagger\right)\nonumber\\
	&={\rm Tr}\,\rho(0)=1.
\end{align}
The open circles of Fig.~\ref{fig:yield} solve Eq.~\eqref{eq:lyapunov} for the cavity and all $N$ emitters, an $(N+1)$-dimensional problem.

Pure dephasing, $L_\varphi^{(m)}=\sqrt{\gamma_\varphi}\,\hat{p}_m^\dagger\hat{p}_m$, does not remove the excitation. Its jumps keep the system in the single-excitation manifold, so $\rho$ becomes the density operator in that manifold, $H_{\rm eff}$ includes the dephasing terms, and Eq.~\eqref{eq:lyapunov} acquires the jump term
\begin{equation}
	\rho(0)=iH_{\rm eff}M-iMH_{\rm eff}^\dagger-\sum_m L_\varphi^{(m)}M L_\varphi^{(m)\dagger},
	\label{eq:lyapunov_deph}
\end{equation}
which is still linear in $M$. Equation~\eqref{eq:phik} is unchanged for the channels that remove the excitation.

\section{Bright mode and dark states}
\label{sm:dark}
\noindent
A single excitation on the emitters, $|\psi\rangle=\sum_m c_m\hat{p}_m^\dagger|{\rm vac}\rangle$ with amplitudes $c_m$, couples to the cavity with amplitude $\langle 1_c|H|\psi\rangle=g\sum_m c_m$. Only the bright state $|B\rangle=\hat{b}^\dagger|{\rm vac}\rangle$, with $\hat{b}=N^{-1/2}\sum_m\hat{p}_m$, couples to the cavity, with strength $\langle 1_c|H|B\rangle=g\sum_m N^{-1/2}=$ $G=g\sqrt{N}$. Every state with $\sum_m c_m=0$ is dark. For example,
\begin{equation}
	|D\rangle = \frac{\hat{p}_1^\dagger - \hat{p}_2^\dagger}{\sqrt{2}}|{\rm vac}\rangle
	\label{eq:dark_state_example}
\end{equation}
has $c_1+c_2=0$ and therefore $\langle 1_c|H|D\rangle=0$. A complete orthonormal set of dark states is given by the discrete Fourier modes
\begin{equation}
	|D_k\rangle = \frac{1}{\sqrt{N}}\sum_{m=1}^{N} e^{2\pi i k m/N}\,\hat{p}_m^\dagger|{\rm vac}\rangle,
	\label{eq:dark_states}
\end{equation}
with $k = 1,\dots,N-1$, while $k=0$ gives $|B\rangle$. This decomposition follows from the star-graph coupling of Eq.~\eqref{eq:H}, with identical emitters of equal dipole orientation.

The incoherent channels discussed in the main text do not couple the bright mode to the dark states. A collective channel, $\sqrt{N\gamma}\,\hat{b}$, adds $-\tfrac{i}{2}N\gamma\,\hat{b}^\dagger\hat{b}$ to $H_{\rm eff}$, which acts only on $|B\rangle$. A local channel with the same rate on every emitter, $\sqrt{\gamma}\,\hat{p}_m$, adds $-\tfrac{i}{2}\gamma\sum_m\hat{p}_m^\dagger\hat{p}_m$, which is proportional to the identity on the emitter states. Starting from $|1_c\rangle$ or $|B\rangle$, the dynamics therefore stays in the space spanned by $|1_c\rangle$ and $|B\rangle$. On resonance, $\omega_c=\omega_0$, and in the frame rotating at $\omega_0$,
\begin{equation}
	H_{\rm eff}=\begin{pmatrix}-i\kappa/2 & G\\ G & -i\gamma_B/2\end{pmatrix}
	\label{eq:Heff2}
\end{equation}
in the basis $\{|1_c\rangle,|B\rangle\}$, where $\gamma_B$ is the total decay rate of the bright mode, $\gamma_B=\gamma_{\rm nr}+N\Gamma$ for collective and $\gamma_B=\gamma_{\rm nr}+\Gamma$ for local conversion. In general, dephasing (Sec.~\ref{sm:dephasing}), disorder and unequal couplings mix $|B\rangle$ with the dark states.

\section{Independent emitters and free space}
\label{sm:freespace}
\noindent
For independent emitters every channel is local, $\sqrt{\Gamma}\,\hat{p}_m$, $\sqrt{\gamma_{\rm r}}\,\hat{p}_m$ and $\sqrt{\gamma_{\rm nr}}\,\hat{p}_m$. In the frame rotating at $\omega_0$ and without the cavity, Eq.~\eqref{eq:Heff_def} on the emitter states reads
\begin{equation}
	H_{\rm eff}=-\tfrac{i}{2}(\Gamma+\gamma_{\rm r}+\gamma_{\rm nr})\sum_m\hat{p}_m^\dagger\hat{p}_m,
\end{equation}
which is proportional to the identity. Equation~\eqref{eq:lyapunov} then gives $M=\rho(0)/(\Gamma+\gamma_{\rm r}+\gamma_{\rm nr})$ for any initial state, and
\begin{equation}
	\phi_{\rm ind}=\Gamma\,{\rm Tr}\Big(\sum_m\hat{p}_m^\dagger\hat{p}_m M\Big)=\frac{\Gamma}{\Gamma+\gamma_{\rm r}+\gamma_{\rm nr}}.
\end{equation}
This holds for any state of the ensemble and any number of excitations it contains, from a single excitation on one emitter to many weakly excited emitters under continuous illumination. The baseline yield is therefore independent of both $N$ and the illumination intensity.

Emitters that share the free-space radiation field radiate collectively through the channel defined by $\sqrt{N\gamma_{\rm r}}\,\hat{b}$. Starting from $|B\rangle$, the bright mode decays exponentially at rate 
\begin{equation}
	\gamma_{\rm tot}=N\gamma_{\rm r}+\gamma_{\rm nr}+\Gamma_{\rm conv},
\end{equation}
with $\Gamma_{\rm conv}=\Gamma$ for local conversion, via $\sqrt{\Gamma}\,\hat{p}_m$, and $\Gamma_{\rm conv}=N\Gamma$ for collective conversion, via $\sqrt{N\Gamma}\,\hat{b}$. Since $H_{\rm eff}|B\rangle=-\tfrac{i}{2}\gamma_{\rm tot}|B\rangle$, Eq.~\eqref{eq:lyapunov} with $\rho(0)=\ketbra{B}{B}$ gives $M=M_{BB}\ketbra{B}{B}$ with
\begin{equation}
	M_{BB}=\frac{1}{\gamma_{\rm tot}},
\end{equation}
and Eq.~\eqref{eq:phik} gives $\phi=\Gamma_{\rm conv}M_{BB}$, because $\sum_m\hat{p}_m^\dagger\hat{p}_m|B\rangle=|B\rangle$ and $N\hat{b}^\dagger\hat{b}|B\rangle=N|B\rangle$. We obtain the yields
\begin{align}
	&\phi_{\rm fs}^{\rm loc}=\frac{\Gamma}{\Gamma+N\gamma_{\rm r}+\gamma_{\rm nr}},
	\label{eq:phi_nocav_loc} \\
    &\phi_{\rm fs}^{\rm col}=\frac{N\Gamma}{N(\Gamma+\gamma_{\rm r})+\gamma_{\rm nr}}.
	\label{eq:phi_nocav_coll}
\end{align}
The local yield falls as $N^{-1}$, suppressed by collective radiation. In the collective yield, conversion and radiation both scale as $N$, and $\phi_{\rm fs}^{\rm col}\to\Gamma/(\Gamma+\gamma_{\rm r})$ at large $N$.

\section{Cavity-coupled emitters, photon start}
\label{sec:methods_cavity_yield}
\noindent
For $|\psi_0\rangle=|1_c\rangle$, we write the elements of $M$ in the basis of Eq.~\eqref{eq:Heff2} as $M_{cc}$, $M_{BB}$ and $M_{cB}=M_{Bc}^*$, the time-integrated populations of the cavity and of the bright mode and their coherence. The three independent elements of Eq.~\eqref{eq:lyapunov} read
\begin{align}
	&\kappa M_{cc}+GX=1,\label{eq:lyap_cc}\\
	&\gamma_B M_{BB}-GX=0,\label{eq:lyap_BB}\\
	&\tfrac{1}{2}(\kappa+\gamma_B)M_{cB}-iG(M_{cc}-M_{BB})=0,\label{eq:lyap_cB}
\end{align}
where $X=2\,{\rm Im}\,M_{cB}$, and $GX$ is the time-integrated flux from the cavity to the bright mode. The sum of Eqs.~\eqref{eq:lyap_cc} and~\eqref{eq:lyap_BB} is the conservation of probability, $\kappa M_{cc}+\gamma_BM_{BB}=1$. The imaginary part of Eq.~\eqref{eq:lyap_cB} and Eq.~\eqref{eq:lyap_BB} give
\begin{align}
	&X=\frac{4G(M_{cc}-M_{BB})}{\kappa+\gamma_B},\\
	&M_{cc}=\frac{\gamma_B(\kappa+\gamma_B)+4G^2}{4G^2}\,M_{BB}.
\end{align}
Inserting $M_{cc}$ into the conservation of probability gives the time-integrated populations
\begin{align}
	&M_{BB}=\frac{4G^2}{(\kappa+\gamma_B)(\kappa\gamma_B+4G^2)},\\
    &M_{cc}=\frac{\gamma_B(\kappa+\gamma_B)+4G^2}{(\kappa+\gamma_B)(\kappa\gamma_B+4G^2)},
	\label{eq:populations}
\end{align}
and the conversion yield factorises as
\begin{align}
	\phi&=\Gamma_{\rm conv}M_{BB}\nonumber\\
	&=\frac{4G^2}{\kappa\gamma_B+4G^2}\cdot\frac{\Gamma_{\rm conv}}{\kappa+\gamma_B}.
	\label{eq:phi_cav_general}
\end{align}
For collective conversion, $\Gamma_{\rm conv}=N\Gamma$ and $\gamma_B=\gamma_{\rm nr}+N\Gamma$,
\begin{equation}
	\phi_{\rm cav}^{\rm col}=\frac{4N\Gamma G^2}{(\kappa+\gamma_{\rm nr}+N\Gamma)[\kappa(\gamma_{\rm nr}+N\Gamma)+4G^2]},
	\label{eq:cav_col_yield}
\end{equation}
which is Eq.~\eqref{eq:phi_cavity} of the main text. For local conversion, $\Gamma_{\rm conv}=\Gamma$ and $\gamma_B=\gamma_{\rm nr}+\Gamma$,
\begin{equation}
	\phi_{\rm cav}^{\rm loc}=\frac{4\Gamma G^2}{(\kappa+\gamma_{\rm nr}+\Gamma)[\kappa(\gamma_{\rm nr}+\Gamma)+4G^2]}.
	\label{eq:cav_loc_yield}
\end{equation}
This yield depends on $N$ only through $G^2$, growing at most linearly, and saturating at $\Gamma/(\Gamma+\kappa+\gamma_{\rm nr})$ for $4G^2\gg\kappa(\gamma_{\rm nr}+\Gamma)$. The cavity removes collective radiation but does not enhance a local channel.

\textit{Plateau.} The first factor of Eq.~\eqref{eq:phi_cavity} compares two rates out of the cavity. We write the no-jump state as $|\widetilde{\psi}(t)\rangle=c_c(t)|1_c\rangle+c_B(t)|B\rangle$. When the bright mode is overdamped, $\gamma_B\gg G$, its amplitude follows the cavity adiabatically,
\begin{align}
	&\dot c_B=-iGc_c-\tfrac{1}{2}\gamma_Bc_B\approx0,\\
	&c_B\approx-\frac{2iG}{\gamma_B}\,c_c,\\
	&\gamma_B|c_B|^2\approx\frac{4G^2}{\gamma_B}|c_c|^2,
\end{align}
so the array absorbs the photon at rate $4G^2/\gamma_B$, in competition with escape at rate $\kappa$. The probability of absorption,
\begin{equation}
	\frac{4G^2/\gamma_B}{\kappa+4G^2/\gamma_B}=\frac{4G^2}{\kappa\gamma_B+4G^2},
\end{equation}
is the first factor of Eq.~\eqref{eq:phi_cav_general}. At large $N$,
\begin{align}
	&\frac{4G^2}{\gamma_B}=\frac{4g^2N}{\gamma_{\rm nr}+N\Gamma}\to\frac{4g^2}{\Gamma},\\
	&\frac{N\Gamma}{\kappa+\gamma_{\rm nr}+N\Gamma}\to1,
\end{align}
so that $\phi_{\rm cav}^{\rm col}\to(4g^2/\Gamma)/(\kappa+4g^2/\Gamma)$, which gives Eq.~\eqref{eq:phi_asymp}.

\textit{Impedance matching.} Increasing $\Gamma$ lowers the plateau of Eq.~\eqref{eq:phi_asymp}, because a bright mode that is drained too quickly can no longer be fed coherently by the cavity. An optimum exists instead in the conversion rate at fixed $N$, when the impedance-matching condition is met
\begin{equation}
	(N\Gamma)^{\star} = \sqrt{(\kappa+\gamma_{\rm nr})\left(\gamma_{\rm nr}+\frac{4G^2}{\kappa}\right)}.
	\label{eq:impedance}
\end{equation}
Slower conversion loses to heating, while faster conversion overdamps the bright mode and blocks its feeding by the cavity. This trade-off between coherent delivery and dissipative loss is a form of impedance matching, as in cavity quantum memories, where a photon is absorbed completely when the absorption by the ensemble matches the coupling of the cavity to the outside~\cite{Afzelius_PRA2010a}, and in charger-mediated quantum batteries, where an intermediate dephasing rate balances coherent energy exchange against quantum Zeno freezing and gives the fastest charging~\cite{Shastri_npjQI2025a}.

At fixed $G$, Eq.~\eqref{eq:cav_col_yield} is proportional to $x/[(A+x)(B+x)]$, with $x=N\Gamma$, $A=\kappa+\gamma_{\rm nr}$ and $B=\gamma_{\rm nr}+4G^2/\kappa$. Its maximum satisfies
\begin{align}
	\frac{d}{dx}\ln\frac{x}{(A+x)(B+x)}&=\frac{1}{x}-\frac{1}{A+x}-\frac{1}{B+x}=0,\\
	(A+x)(B+x)&=x(A+x)+x(B+x),\\
	x^2&=AB,
\end{align}
which is Eq.~\eqref{eq:impedance}.

\textit{Threshold.} Equation~\eqref{eq:phi_asymp} exceeds $\phi_{\rm ind}$ when
\begin{align}
	\frac{4g^2}{4g^2+\kappa\Gamma}&>\frac{\Gamma}{\Gamma+\gamma_{\rm r}+\gamma_{\rm nr}},\\
	4g^2(\gamma_{\rm r}+\gamma_{\rm nr})&>\kappa\Gamma^2,
\end{align}
which is Eq.~\eqref{eq:coop_threshold}.

\section{Scaling exponent}
\label{sm:alpha}
\noindent
With $G^2=g^2N$ and $\kappa(\gamma_{\rm nr}+N\Gamma)+4g^2N=\kappa\gamma_{\rm nr}+N(\kappa\Gamma+4g^2)$, Eq.~\eqref{eq:cav_col_yield} reads
\begin{equation}
	\phi_{\rm cav}^{\rm col}=\frac{4g^2\Gamma N^2}{(\kappa+\gamma_{\rm nr}+N\Gamma)[\kappa\gamma_{\rm nr}+N(\kappa\Gamma+4g^2)]}.
\end{equation}
Taking the logarithm and differentiating with respect to $\ln N$, the numerator contributes 2 and each factor of the denominator subtracts one term,
\begin{equation}
	\alpha(N)=\frac{d\ln\phi_{\rm cav}^{\rm col}}{d\ln N}=2-f_{\rm conv}-f_{\rm abs},
	\label{eq:alpha_cav}
\end{equation}
with
\begin{align}
	f_{\rm conv}&=\frac{N\Gamma}{\kappa+\gamma_{\rm nr}+N\Gamma},\\
	f_{\rm abs}&=\frac{N(\kappa\Gamma+4g^2)}{\kappa\gamma_{\rm nr}+N(\kappa\Gamma+4g^2)},
\end{align}
where $f_{\rm abs}$ and $f_{\rm conv}$ come from the absorption and conversion factors of Eq.~\eqref{eq:phi_cavity}. Both rise from 0 to 1, reaching one half at
\begin{align}
	N_1 &= \frac{\kappa\gamma_{\rm nr}}{\kappa\Gamma + 4g^2},\\
	N_2 &= \frac{\kappa+\gamma_{\rm nr}}{\Gamma},
	\label{eq:N1N2}
\end{align}
respectively. Their ratio,
\begin{align}
	\frac{N_2}{N_1}&=\frac{(\kappa+\gamma_{\rm nr})(\kappa\Gamma+4g^2)}{\kappa\gamma_{\rm nr}\Gamma}\nonumber\\
	&=1+\frac{\kappa}{\gamma_{\rm nr}}+\frac{4g^2(\kappa+\gamma_{\rm nr})}{\kappa\gamma_{\rm nr}\Gamma},
	\label{eq:N2overN1}
\end{align}
exceeds one, so $N_1<N_2$ and $\alpha$ decreases monotonically from 2 to 0 through four regimes.
\begin{itemize}
	\item  $\alpha\approx2$ for $N\ll N_1$, where both factors grow linearly, so that $\phi\approx4g^2N/(\kappa\gamma_{\rm nr})  \times N\Gamma/(\kappa+\gamma_{\rm nr})$.
	\item $\alpha\in(1,2)$ near $N_1$, the strong-coupling crossover. In terms of the collective cooperativity of the main text, $C_N=4G^2/(\kappa\gamma_B)$ with $\gamma_B=\gamma_{\rm nr}+N\Gamma$, the absorption term splits as
	\begin{equation}
		f_{\rm abs}=\frac{C_N}{1+C_N}+\frac{\kappa\Gamma N}{\kappa\gamma_B+4G^2}.
	\end{equation}
	For $N\Gamma\ll\gamma_{\rm nr}$, the second term and $f_{\rm conv}$ are both smaller than $N\Gamma/\gamma_{\rm nr}$, so $\alpha\approx2-C_N/(1+C_N)$, and $\alpha=3/2$ at $C_N=1$.
	\item $\alpha\approx1$ for $N_1\ll N\ll N_2$, where absorption is saturated and $\phi\approx N\Gamma/(\kappa+\gamma_{\rm nr})$. This regime is wide when $N_2/N_1\gg1$, that is, when the last term of Eq.~\eqref{eq:N2overN1} is large.
	\item $\alpha\approx0$ for $N\gg N_2$, where conversion dominates the losses of the bright mode, and $\phi$ reaches the plateau of Eq.~\eqref{eq:phi_asymp}.
\end{itemize}

\begin{table*}[t]
	\centering
	\renewcommand{\arraystretch}{1.6}
	\setlength{\tabcolsep}{6pt}
	\begin{tabular}{@{}llccc@{}}
		\toprule
		\textbf{Configuration} & \textbf{Conversion} $L_\Gamma$ & \textbf{Initial state} & \textbf{Yield } $\phi(N)$ & \textbf{Limit } $N\to\infty$ \\
		\midrule
		\multicolumn{5}{@{}l}{\textit{\textbf{No cavity}}} \\[2pt]
		Independent emitters  & $\sqrt{\Gamma}\,\hat{p}_m$ & any & $\dfrac{\Gamma}{\Gamma+\gamma_{\rm r}+\gamma_{\rm nr}}$ & $\dfrac{\Gamma}{\Gamma+\gamma_{\rm r}+\gamma_{\rm nr}}$ \\[8pt]
		Local conversion      & $\sqrt{\Gamma}\,\hat{p}_m$ & $|B\rangle$   & $\dfrac{\Gamma}{\Gamma+N\gamma_{\rm r}+\gamma_{\rm nr}}$ & $0$ \\[8pt]
		Collective conversion & $\sqrt{N\Gamma}\,\hat{b}$  & $|B\rangle$   & $\dfrac{N\Gamma}{N(\Gamma+\gamma_{\rm r})+\gamma_{\rm nr}}$ & $\dfrac{\Gamma}{\Gamma+\gamma_{\rm r}}$ \\
		\addlinespace[8pt]
		\multicolumn{5}{@{}l}{\textit{\textbf{Cavity, photon start}}} \\[2pt]
		Local conversion      & $\sqrt{\Gamma}\,\hat{p}_m$ & $|1_c\rangle$ & $\dfrac{4\Gamma G^2}{(\kappa+\Gamma+\gamma_{\rm nr})[\kappa(\Gamma+\gamma_{\rm nr})+4G^2]}$ & $\dfrac{\Gamma}{\Gamma+\kappa+\gamma_{\rm nr}}$ \\[8pt]
		Collective conversion & $\sqrt{N\Gamma}\,\hat{b}$  & $|1_c\rangle$ & $\dfrac{4N\Gamma G^2}{(\kappa+\gamma_{\rm nr}+N\Gamma)[\kappa(\gamma_{\rm nr}+N\Gamma)+4G^2]}$ & $\dfrac{4g^2}{\kappa\Gamma+4g^2}$ \\
		\addlinespace[8pt]
		\multicolumn{5}{@{}l}{\textit{\textbf{Cavity, polariton start}}} \\[2pt]
		Local conversion      & $\sqrt{\Gamma}\,\hat{p}_m$ & $|P_\pm\rangle$ & $\dfrac{\Gamma[\kappa(\Gamma+\gamma_{\rm nr}+\kappa)+8G^2]}{2(\Gamma+\gamma_{\rm nr}+\kappa)[\kappa(\Gamma+\gamma_{\rm nr})+4G^2]}$ & $\dfrac{\Gamma}{\Gamma+\kappa+\gamma_{\rm nr}}$ \\[8pt]
		Collective conversion & $\sqrt{N\Gamma}\,\hat{b}$  & $|P_\pm\rangle$ & $\dfrac{N\Gamma[8G^2+\kappa(\gamma_{\rm nr}+\kappa+N\Gamma)]}{2(\gamma_{\rm nr}+\kappa+N\Gamma)[\kappa(\gamma_{\rm nr}+N\Gamma)+4G^2]}$ & $\dfrac{\kappa\Gamma+8g^2}{2(\kappa\Gamma+4g^2)}$ \\
		\addlinespace[8pt]
		\multicolumn{5}{@{}l}{\textit{\textbf{Cavity, steady drive at $\delta=0$, per dissipated photon}}} \\[2pt]
		Local conversion      & $\sqrt{\Gamma}\,\hat{p}_m$ & drive & $\dfrac{4\Gamma G^2}{(\Gamma+\gamma_{\rm nr})[\kappa_0(\Gamma+\gamma_{\rm nr})+4G^2]}$ & $\dfrac{\Gamma}{\Gamma+\gamma_{\rm nr}}$ \\[8pt]
		Collective conversion & $\sqrt{N\Gamma}\,\hat{b}$  & drive & $\dfrac{4N\Gamma G^2}{(\gamma_{\rm nr}+N\Gamma)[\kappa_0(\gamma_{\rm nr}+N\Gamma)+4G^2]}$ & $\dfrac{4g^2}{\kappa_0\Gamma+4g^2}$ \\
		\bottomrule
	\end{tabular}
	\caption{\textbf{Conversion yields.}  Conversion yields for the considered configurations, assuming $G=g\sqrt{N}$. All configurations include local non-radiative loss, $\sqrt{\gamma_{\rm nr}}\,\hat{p}_m$. Without the cavity, radiation is local for independent emitters, $\sqrt{\gamma_{\rm r}}\,\hat{p}_m$, and collective otherwise, $\sqrt{N\gamma_{\rm r}}\,\hat{b}$. With the cavity, light leaves through $\sqrt{\kappa}\,\hat{a}$, and for the steady drive only the mirror absorption $\kappa_0$ counts as a loss (Sec.~\ref{sm:driven}). At large $N$, the collective photon-start yield exceeds that of independent emitters when $4g^2/(\kappa\Gamma)>\Gamma/(\gamma_{\rm r}+\gamma_{\rm nr})$.}
	\label{tab:yields}
\end{table*}

\section{Bright-mode and polariton starts}
\label{sm:initial-states}
\begin{figure}
	\centering
	\includegraphics[width=\columnwidth]{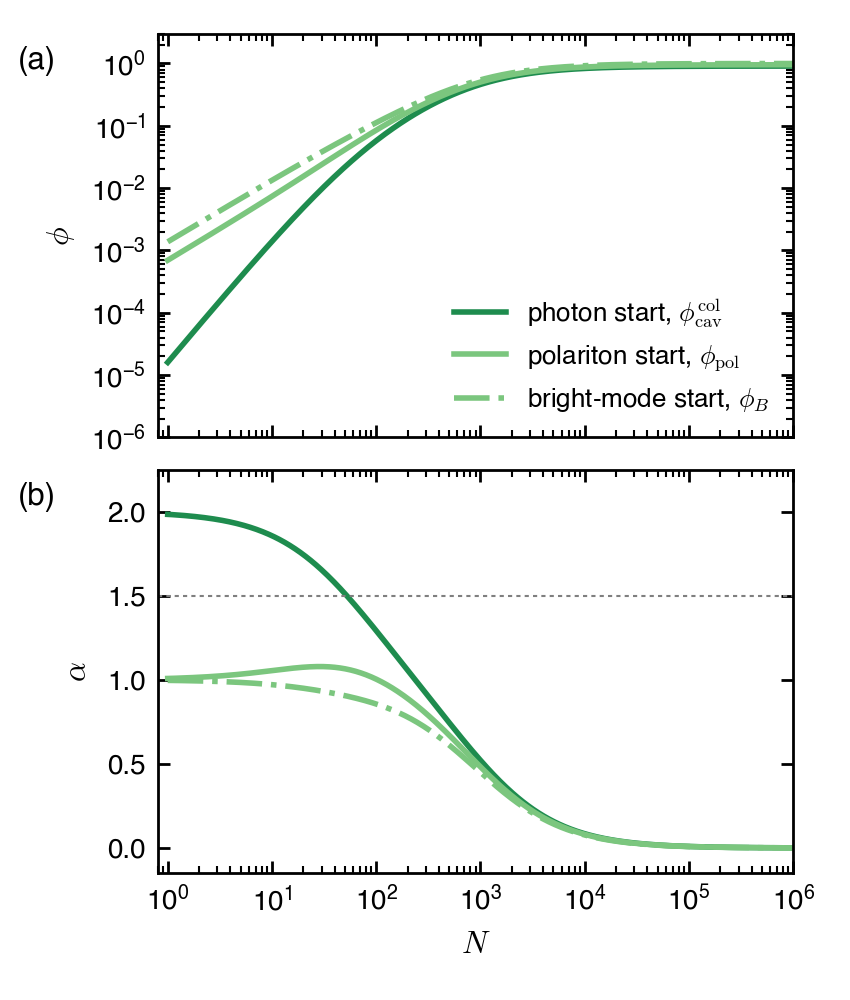}
	\caption{\textbf{Yield and scaling exponent for the different initial states.} (a)~Collective conversion yield in the cavity for a photon, polariton and bright-mode start, and (b)~its exponent $\alpha$, with the parameters of Fig.~\ref{fig:yield}. Only the photon start reaches $\alpha=2$.}
	\label{fig:initial_states}
\end{figure}

\noindent
Let us now consider the case where the state is initialised in the bright mode or in a polariton. On resonance the polaritons are $|P_\pm\rangle=(|1_c\rangle\pm|B\rangle)/\sqrt{2}$, split by $2G$. For $\rho(0)=\ketbra{B}{B}$, the right-hand sides of Eqs.~\eqref{eq:lyap_cc} and~\eqref{eq:lyap_BB} become 0 and 1, while Eq.~\eqref{eq:lyap_cB} and the conservation of probability are unchanged. Equation~\eqref{eq:lyap_cc} then gives
\begin{align}
	&\kappa M_{cc}=-GX=-\frac{4G^2(M_{cc}-M_{BB})}{\kappa+\gamma_B},\\
	&M_{cc}=\frac{4G^2}{\kappa(\kappa+\gamma_B)+4G^2}\,M_{BB},\\
	&M_{BB}=\frac{\kappa(\kappa+\gamma_B)+4G^2}{(\kappa+\gamma_B)(\kappa\gamma_B+4G^2)},
\end{align}
where the last line follows from $\kappa M_{cc}+\gamma_BM_{BB}=1$. The yield is $\phi_B=\Gamma_{\rm conv}M_{BB}$, which for collective conversion reads
\begin{equation}
	\phi_B=\frac{N\Gamma[\kappa(\kappa+\gamma_B)+4G^2]}{(\kappa+\gamma_B)(\kappa\gamma_B+4G^2)}.
	\label{eq:phi_bright}
\end{equation}
For a polariton, $\rho(0)=\tfrac12(\ketbra{1_c}{1_c}+\ketbra{B}{B})\pm\tfrac12(\ketbra{1_c}{B}+\ketbra{B}{1_c})$. Since Eq.~\eqref{eq:lyapunov} is linear, $M$ is the sum of the solutions for the three terms. For the cross term, the right-hand sides of Eqs.~\eqref{eq:lyap_cc} and~\eqref{eq:lyap_BB} vanish, and the conservation of probability becomes $\kappa M_{cc}+\gamma_BM_{BB}=0$. Combining it with Eq.~\eqref{eq:lyap_BB} and the imaginary part of Eq.~\eqref{eq:lyap_cB}, which are unchanged, gives
\begin{align}
	\gamma_BM_{BB}&=\frac{4G^2(M_{cc}-M_{BB})}{\kappa+\gamma_B}=-\frac{4G^2}{\kappa}M_{BB},
\end{align}
so that $M_{BB}=0$. The cross term does not contribute, and both polaritons give the average of the photon and bright-mode starts,
\begin{equation}
	\phi_{\rm pol}=\frac{\Gamma_{\rm conv}[8G^2+\kappa(\kappa+\gamma_B)]}{2(\kappa+\gamma_B)(\kappa\gamma_B+4G^2)},
	\label{eq:phi_pol}
\end{equation}
for both local and collective conversion (Table~\ref{tab:yields}). For collective conversion, this exceeds the photon-start yield by the factor $1+\kappa(\kappa+\gamma_B)/(8G^2)$, which tends to one at strong coupling, and saturates at $(\kappa\Gamma+8g^2)/[2(\kappa\Gamma+4g^2)]$ at large $N$. At small $N$, $G^2=g^2N\to0$ and $\gamma_B\to\gamma_{\rm nr}$, so that
\begin{align}
	\phi_B&\approx\frac{N\Gamma}{\gamma_{\rm nr}},\\
	\phi_{\rm pol}&\approx\frac{\phi_B}{2}\approx\frac{N\Gamma}{2\gamma_{\rm nr}},
\end{align}
because $\phi_{\rm cav}^{\rm col}\propto N^2$ is negligible. Part of the excitation starts in the bright mode and bypasses absorption, so $\phi_B$ and $\phi_{\rm pol}$ grow linearly, with $\alpha\approx1$, and only the photon start reaches $\alpha=2$ (Fig.~\ref{fig:initial_states}). Although the experiment pumps the polaritons, it is described by the steady drive of Sec.~\ref{sm:driven}, which retains the absorption step.

\section{Driven yield per dissipated photon}
\label{sm:driven}
\noindent
Equation~\eqref{eq:phi_drive} follows from the same model as Eq.~\eqref{eq:cav_col_yield}, evaluated for the measured quantity. In both cases the yield is the converted flux divided by the flux that is counted,
\begin{equation}
	\phi=\frac{N_{\rm coh}\Gamma\,P_B}{\kappa_{\rm L}P_c+\gamma_BP_B}=\frac{N_{\rm coh}\Gamma}{\gamma_B+\kappa_{\rm L}\,P_c/P_B},
	\label{eq:yield_ratio}
\end{equation}
where $P_c$ and $P_B$ are the populations of the cavity and of the bright mode, $\gamma_B=\gamma_{\rm nr}+N_{\rm coh}\Gamma$, $\kappa_{\rm L}$ is the part of the cavity loss that counts as a failure, and $N_{\rm coh}=N$ for the ideal array.

\textit{Photon start.} With $P_c=M_{cc}$ and $P_B=M_{BB}$ from Eq.~\eqref{eq:populations}, every photon that leaves the cavity is lost, $\kappa_{\rm L}=\kappa$, and Eq.~\eqref{eq:yield_ratio} reproduces Eq.~\eqref{eq:cav_col_yield},
\begin{align}
	\phi&=\frac{N\Gamma}{\gamma_B+\kappa M_{cc}/M_{BB}}\nonumber\\
	&=\frac{4G^2N\Gamma}{4G^2\gamma_B+\kappa[\gamma_B(\kappa+\gamma_B)+4G^2]}\nonumber\\
	&=\frac{4G^2N\Gamma}{(\kappa+\gamma_B)(\kappa\gamma_B+4G^2)}.
\end{align}

\textit{Steady drive.} In the experiment the cavity is driven at frequency $\omega$ through its coupling $\kappa_{\rm ex}$ to the incoming light, and its total loss is $\kappa=\kappa_{\rm ex}+\kappa_0$, where $\kappa_0$ is parasitic absorption in the Au mirror. For an incoming field of amplitude $a_{\rm in}$, the amplitudes of the cavity and of the bright mode, whose frequency is $\Omega_B$, obey
\begin{align}
	\frac{d\langle\hat{a}\rangle}{dt}&=-\Big(i\omega_c+\frac{\kappa}{2}\Big)\langle\hat{a}\rangle-iG\langle\hat{b}\rangle+\sqrt{\kappa_{\rm ex}}\,a_{\rm in}e^{-i\omega t},\\
	\frac{d\langle\hat{b}\rangle}{dt}&=-\Big(i\Omega_B+\frac{\gamma_B}{2}\Big)\langle\hat{b}\rangle-iG\langle\hat{a}\rangle.
\end{align}
For a pump that is long compared with the inverse linewidths, the amplitudes reach a steady state that oscillates at $\omega$. The drive acts only on the cavity, so the equation of motion of the bright mode gives
\begin{equation}
    G\langle\hat{a}\rangle+(\Omega_B-\omega-i\gamma_B/2)\langle\hat{b}\rangle=0,
\end{equation}
and the population ratio
\begin{equation}
    \frac{P_c}{P_B}=\frac{|\langle\hat{a}\rangle|^2}{|\langle\hat{b}\rangle|^2}=\frac{\delta^2+\gamma_B^2/4}{G^2},
\end{equation}
with $\delta=\omega-\Omega_B$ the detuning of the pump from the bright mode. The AQE of the ATA structures is normalised by the absorbance $1-R$, that is, per photon dissipated in the structure. Light that leaves the cavity through the input side is reflected and is not counted, so only mirror absorption is a loss channel, $\kappa_{\rm L}=\kappa_0$. Equation~\eqref{eq:yield_ratio} then gives Eq.~\eqref{eq:phi_drive}, which we rewrite as
\begin{align}
	\phi_{\rm ATA}&=\frac{N_{\rm coh}\Gamma\,G^2}{\kappa_0\left(\delta^2+\gamma_B^2/4\right)+\gamma_BG^2}\nonumber\\
	&=\frac{4G^2}{\kappa_0\gamma_B\left(1+4\delta^2/\gamma_B^2\right)+4G^2}\cdot\frac{N_{\rm coh}\Gamma}{\gamma_B},
	\label{eq:phi_drive_factor}
\end{align}
independent of $\kappa_{\rm ex}$. The first factor is the share of the dissipated light absorbed by the array rather than by the mirror. The second is the share of the dissipation of the array that goes into conversion. The cavity loss enters Eq.~\eqref{eq:cav_col_yield} twice, as $\kappa_{\rm L}$ and through $P_c/P_B$, while only $\kappa_0$ competes in Eq.~\eqref{eq:phi_drive_factor}. For $N_{\rm coh}=N$, $\delta=0$ and a fixed single-emitter coupling, $G^2=g^2N$, Eq.~\eqref{eq:phi_drive_factor} reads
\begin{equation}
	\phi_{\rm ATA}=\frac{4g^2\Gamma N^2}{(\gamma_{\rm nr}+N\Gamma)[\kappa_0\gamma_{\rm nr}+N(\kappa_0\Gamma+4g^2)]},
\end{equation}
and its exponent, obtained as in Sec.~\ref{sm:alpha}, is
\begin{equation}
	\alpha = 2-\frac{N\Gamma}{\gamma_{\rm nr}+N\Gamma}-\frac{N(\kappa_0\Gamma+4g^2)}{\kappa_0\gamma_{\rm nr}+N(\kappa_0\Gamma+4g^2)},
	\label{eq:alpha_drive}
\end{equation}
which has the form of Eq.~\eqref{eq:alpha_cav} with $\kappa$ replaced by $\kappa_0$ in the absorption term and removed from the conversion term. It flows from 2 to 0 in the same way. In the measured samples $g$ grows with density, so the measured exponent cannot be compared directly with Eq.~\eqref{eq:alpha_drive}. Section~\ref{sec:local_vs_collective} therefore separates the optical factor from the transfer rate. The values of $\kappa_0$, $G$ and $\gamma_B$ of the measured samples are obtained in Secs.~\ref{sec:fit_ata} and~\ref{sec:local_vs_collective}.

\begin{figure*}
	\centering
	\includegraphics[width=75mm]{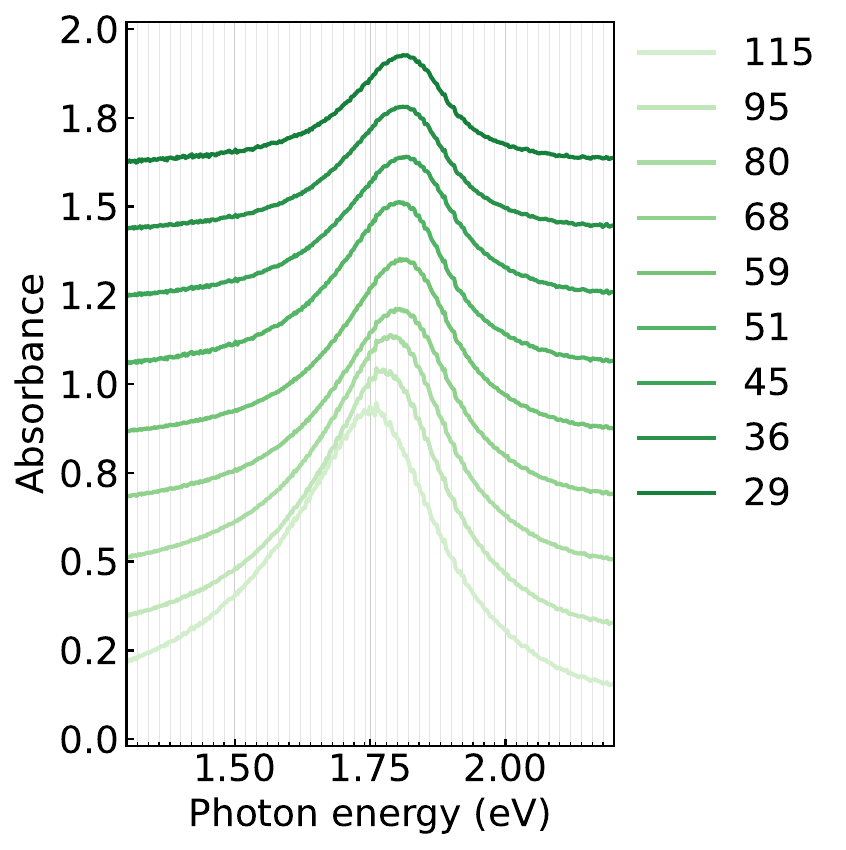}
	\includegraphics[width=75mm]{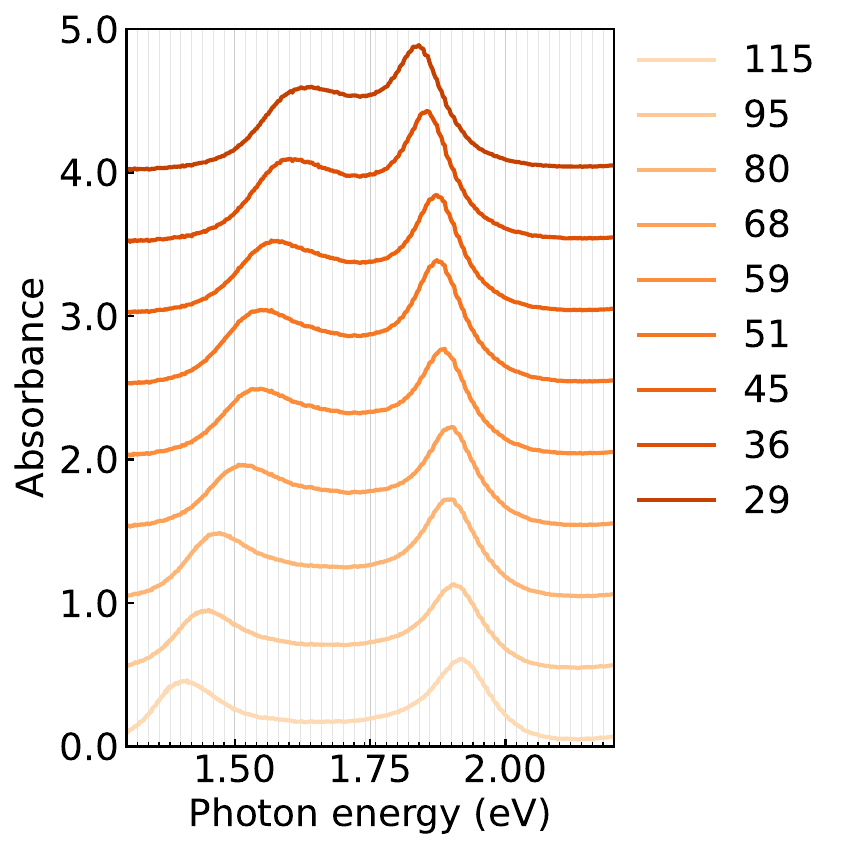}
	\caption{\textbf{Steady-state absorption spectra.} Steady-state absorption spectra of the AT structures without cavity (left) and of the ATA structures with cavity (right). The legend gives the particle number density in $\mu$m$^{-2}$. The spectra are offset vertically for clarity. Data from Ref.~\cite{Liu_AN2023a}.}
	\label{fig:steadystatespectra}
\end{figure*}

\section{Pure dephasing}
\label{sm:dephasing}
\noindent
Local dephasing, $L_\varphi^{(m)}=\sqrt{\gamma_\varphi}\,\hat{p}_m^\dagger\hat{p}_m$, randomises the relative phases of the emitters. A jump on emitter $m$ projects the excitation onto that emitter,
\begin{equation}
	L_\varphi^{(m)}|B\rangle=\sqrt{\gamma_\varphi/N}\;\hat{p}_m^\dagger|{\rm vac}\rangle.
\end{equation}
Summing over $m$, jumps occur at the total rate $\sum_m\|L_\varphi^{(m)}|B\rangle\|^2=\gamma_\varphi$, and a fraction $|\langle B|\hat{p}_m^\dagger|{\rm vac}\rangle|^2=1/N$ of the excitation returns to $|B\rangle$. Dephasing therefore scatters the bright mode into the dark states at rate $\gamma_\varphi(1-1/N)$. By the same argument, a dark state $\sum_mc_m\hat{p}_m^\dagger|{\rm vac}\rangle$ returns to $|B\rangle$ at rate $\gamma_\varphi\sum_m|c_m|^2/N=\gamma_\varphi/N$. Collective conversion does not act on the dark states. Up to corrections of order $1/N$, the solution of Eq.~\eqref{eq:lyapunov_deph} therefore adds $\gamma_\varphi$ to the width of the bright mode, $\gamma_B\to\gamma_B+\gamma_\varphi$, in Eqs.~\eqref{eq:cav_col_yield} and~\eqref{eq:phi_drive_factor}. The collective enhancement survives for $\gamma_\varphi\ll\gamma_B$.

Static disorder of the site energies mixes the bright mode with dark states of similar energy. In the nearest-neighbour model of Sec.~\ref{sec:experiment_SI}, the sum of cosines in Eq.~\eqref{eq:dispersion} ranges from $-3/2$ to $3$, so with $J<0$ the plasmon band spans $\Omega_B\le\Omega_{\rm pl}(\mathbf{k})\le\Omega_B+9|J|$, and the bright mode lies at its lower edge. In the ATA structures the lower polariton lies 160 to 330~meV below this band, well beyond its half width of 65 to 100~meV, while the upper polariton lies 25 to 75~meV above it, comparable to its half width of 50 to 65~meV.

\section{Experimental data}
\label{sec:experiment_SI}
All data considered in this work were obtained from Ref.~\cite{Liu_AN2023a}. The steady-state absorption spectra are shown in Fig.~\ref{fig:steadystatespectra}. The AT structures show a single absorption band at all particle number densities (PND). The ATA structures always show a doublet, the upper and lower polaritons formed by strong light--matter coupling.

\begin{figure*}
	\centering
	\includegraphics[width=\textwidth]{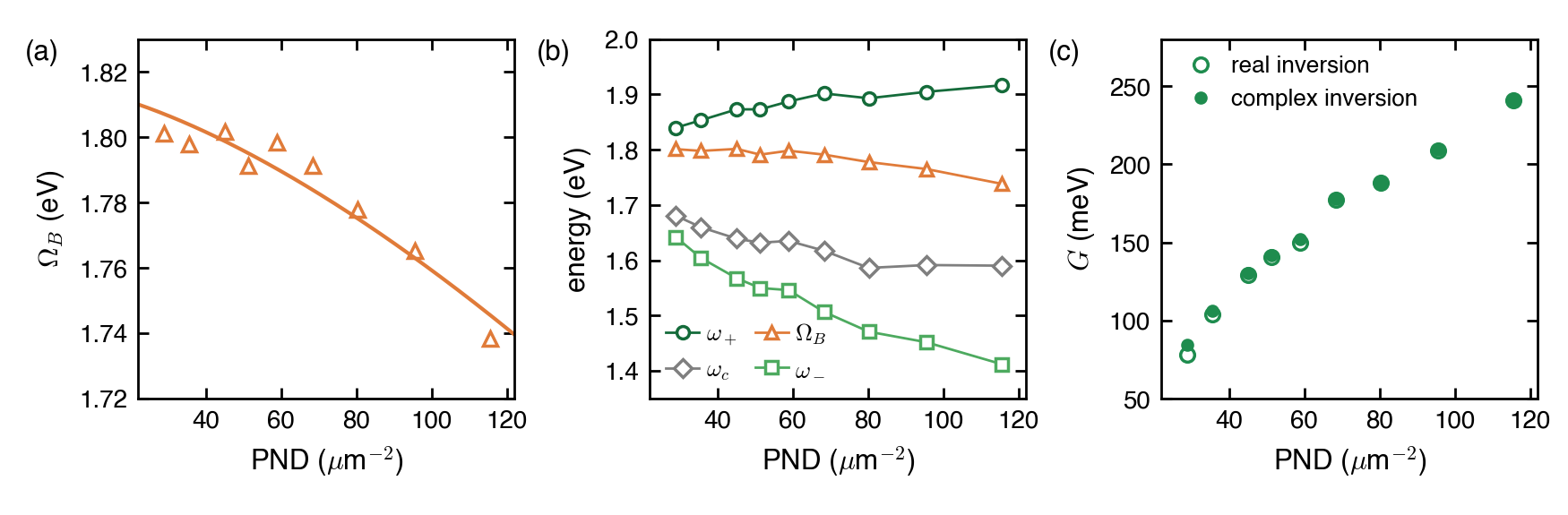}
	\caption{\textbf{Density-dependent resonance and coupling.} (a)~Bright-mode energy $\Omega_B$ of the AT structures, with the fit of Eq.~\eqref{eq:brightmodeshift}, $\omega_0 = 1.816\pm0.004$~eV and $|J_0| = 0.0117\pm0.0013$~meV\,$\mu$m$^3$. (b)~Polariton energies $\omega_\pm$, bright-mode energy $\Omega_B$ and cavity frequency $\omega_c$ from the real inversion. (c)~Collective coupling $G$ from the real inversion (open) and from the complex inversion of Eq.~\eqref{eq:complex_inversion} (filled), which is used in the main text. Linewidths are shown in Fig.~\ref{fig:spectral}.}
	\label{fig:hybridisation}
\end{figure*}

\subsection{Density-dependent resonance shift}
\noindent
The AT absorption maximum redshifts as PND increases, because neighbouring nanoparticles couple through their near fields. On the hexagonal (triangular) lattice each particle has six nearest neighbours, at spacing $a$. Each particle occupies a unit cell of area $\sqrt{3}a^2/2$, so that ${\rm PND}=2/(\sqrt{3}a^2)$ and
\begin{equation}
    a=\sqrt{2/(\sqrt{3}\,{\rm PND})},
\end{equation}
between 100 and 200~nm here. This shift must be separated from the light--matter coupling, so we take the bright-mode frequency $\Omega_B$ at each density from the AT peak.

Nearest-neighbour coupling is described by
\begin{equation}
	H = \omega_0\sum_{m=1}^{N}\hat{p}_m^\dagger\hat{p}_m + J\sum_{\langle m,n\rangle}\left(\hat{p}_m^\dagger\hat{p}_n + \hat{p}_n^\dagger\hat{p}_m\right),
	\label{eq:nearneighbor}
\end{equation}
where $J$ is the hopping energy and the sum runs over nearest-neighbour pairs. The Bloch states $|\mathbf{k}\rangle=N^{-1/2}\sum_me^{i\mathbf{k}\cdot\mathbf{r}_m}\hat{p}_m^\dagger|{\rm vac}\rangle$, with $\mathbf{r}_m$ the position of particle $m$, are eigenstates of Eq.~\eqref{eq:nearneighbor}, because the hopping term multiplies each of them by the sum of $e^{i\mathbf{k}\cdot\boldsymbol{\delta}}$ over the six neighbour vectors $\boldsymbol{\delta}$. These are $\pm\boldsymbol{\delta}_j$, with $\boldsymbol{\delta}_1 = a(1,0)$, $\boldsymbol{\delta}_2 = a(1/2,\sqrt{3}/2)$ and $\boldsymbol{\delta}_3 = a(-1/2,\sqrt{3}/2)$, so the plasmon dispersion is
\begin{equation}
	\Omega_{\mathrm{pl}}(\mathbf{k}) = \omega_0 + 2J \sum_{j=1}^{3}\cos(\mathbf{k}\cdot \boldsymbol{\delta}_j).
	\label{eq:dispersion}
\end{equation}
The bright mode is the $\mathbf{k}=0$ state, $|B\rangle=N^{-1/2}\sum_m\hat{p}_m^\dagger|{\rm vac}\rangle$, with energy
\begin{equation}
	\Omega_B = \Omega_{\mathrm{pl}}(\mathbf{0}) = \omega_0 + 6J.
	\label{eq:brightmodetheory}
\end{equation}
Since $|B\rangle$ is an eigenstate of Eq.~\eqref{eq:nearneighbor}, a uniform $J$ only shifts the bright-mode frequency from $\omega_0$ to $\Omega_B$, and the reduction of Sec.~\ref{sm:dark} holds with $\omega_0$ replaced by $\Omega_B$. The near-field coupling scales as $J=-|J_0|/a^3$, where the sign gives the observed redshift, so
\begin{equation}
	\Omega_B = \omega_0 - \frac{6|J_0|}{a^3}.
	\label{eq:brightmodeshift}
\end{equation}
A fit of Eq.~\eqref{eq:brightmodeshift} to the AT peak energies gives $\omega_0 = 1.816\pm0.004$~eV and $|J_0| = 0.0117\pm0.0013$~meV\,$\mu$m$^3$, as in Fig.~\ref{fig:hybridisation}~(a), with shifts $6|J|$ from 9 to 70~meV. This Hermitian model describes the shift but not the growth of the AT linewidth, from 290 to 360~THz.

\subsection{Cavity frequency and coupling}
\label{sec:fit_ata}
\noindent
The polaritons are the eigenstates of the cavity coupled to the bright mode, described by
\begin{equation}
	H_{\rm eff}=\begin{pmatrix}\omega_c-i\kappa/2 & G\\ G & \Omega_B-i\gamma_B/2\end{pmatrix}
	\label{eq:Heff_exp}
\end{equation}
in the basis $\{|1_c\rangle,|B\rangle\}$. Neglecting the linewidths, the measured polariton energies $\omega_\pm$ are the eigenvalues of its real part,
\begin{equation}
	\omega_\pm=\frac{\omega_c+\Omega_B}{2}\pm\frac{1}{2}\sqrt{\Delta^2+4G^2},
\end{equation}
with $\Delta=\omega_c-\Omega_B$. Their sum and difference,
\begin{align}
	&\omega_++\omega_-=\omega_c+\Omega_B,\\
	&(\omega_+-\omega_-)^2=\Delta^2+4G^2,
\end{align}
give
\begin{align}
    &\omega_c=\omega_++\omega_--\Omega_B,\\
    &G=\tfrac12\sqrt{(\omega_+-\omega_-)^2-(\omega_c-\Omega_B)^2},
\end{align}
at each density, as shown in Fig.~\ref{fig:hybridisation}~(b) and~(c). The coupling grows faster than $\sqrt{N}$ and $\omega_c$ redshifts with density, because the nanodisk array forms the top mirror of the cavity. We therefore obtain $\omega_c$ and $G$ at each density instead of holding them fixed. The complex inversion below also includes the linewidths and is used in the main text.

\paragraph*{Linewidths, cavity loss and cooperativity.}
\label{sec:spectral_params}
To obtain the cavity loss $\kappa$, the bright-mode width $\gamma_B$ and the cooperativity $C=4G^2/(\kappa\gamma_B)$ entering Eq.~\eqref{eq:phi_drive}, we treat the polariton peaks as the complex eigenvalues $\lambda_\pm=\omega_\pm-i\gamma_\pm/2$ of Eq.~\eqref{eq:Heff_exp}, where $\gamma_\pm$ are the full widths at half maximum of the ATA peaks and $\Omega_B$ is the AT peak energy at the same density. Each width is twice the half width on the outer side of its peak, which avoids the overlap of the two polaritons. The eigenvalues of Eq.~\eqref{eq:Heff_exp} are
\begin{align}
	\lambda_\pm&=\frac{\omega_c+\Omega_B}{2}-\frac{i(\kappa+\gamma_B)}{4}\nonumber\\
	&\quad\pm\frac{1}{2}\sqrt{\left[\Delta-\tfrac{i}{2}(\kappa-\gamma_B)\right]^2+4G^2}.
\end{align}

Their sum and the square of their difference,
\begin{align}
	\lambda_++\lambda_-&=\omega_c+\Omega_B-\tfrac{i}{2}(\kappa+\gamma_B),\\
	(\lambda_+-\lambda_-)^2&=\left[\Delta-\tfrac{i}{2}(\kappa-\gamma_B)\right]^2+4G^2,
\end{align}
are compared with the measured values, $\lambda_++\lambda_-=\omega_++\omega_--\tfrac{i}{2}(\gamma_++\gamma_-)$ and
\begin{align}
	(\lambda_+-\lambda_-)^2&=(\omega_+-\omega_-)^2-\tfrac14(\gamma_+-\gamma_-)^2\nonumber\\
	&\quad-i(\omega_+-\omega_-)(\gamma_+-\gamma_-).
\end{align}
The sum fixes $\omega_c$ and $\kappa+\gamma_B$. The imaginary part of the squared difference fixes $\kappa-\gamma_B$ and its real part fixes $G$,
\begin{equation}
	\begin{aligned}
	\Delta&=\omega_++\omega_--2\Omega_B,\\
	\kappa+\gamma_B&=\gamma_++\gamma_-,\\
	\kappa-\gamma_B&=(\omega_+-\omega_-)(\gamma_+-\gamma_-)/\Delta,\\
	4G^2&=(\omega_+-\omega_-)^2-\Delta^2\\
	&\quad-[(\gamma_+-\gamma_-)^2-(\kappa-\gamma_B)^2]/4.
	\end{aligned}
	\label{eq:complex_inversion}
\end{equation}
\begin{figure}[t]
	\centering
	\includegraphics[width=\columnwidth]{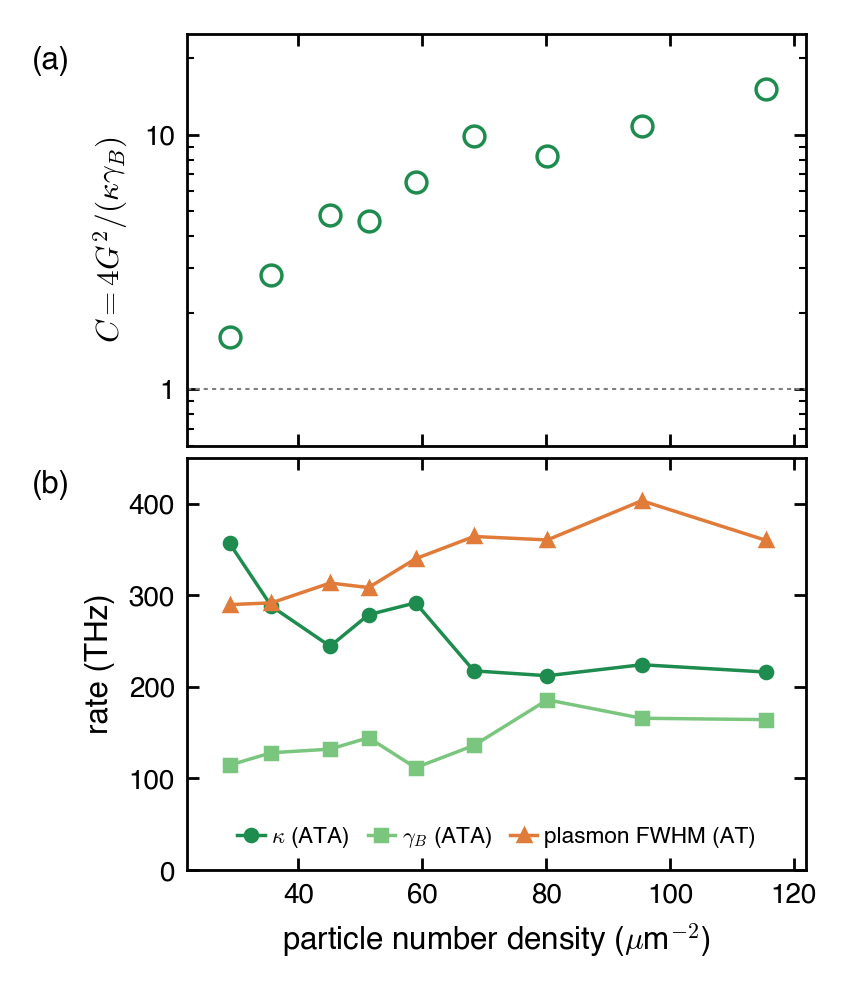}
	\caption{\textbf{Parameters from the measured spectra.} (a)~Cooperativity $C=4G^2/(\kappa\gamma_B)$ of the ATA structures, from Eq.~\eqref{eq:complex_inversion}. The dotted line marks $C=1$. (b)~Cavity loss $\kappa$ and bright-mode width $\gamma_B$ of the ATA structures, and linewidth of the AT plasmon.}
	\label{fig:spectral}
\end{figure}
\begin{figure*}[t]
	\centering
	\includegraphics[width=\textwidth]{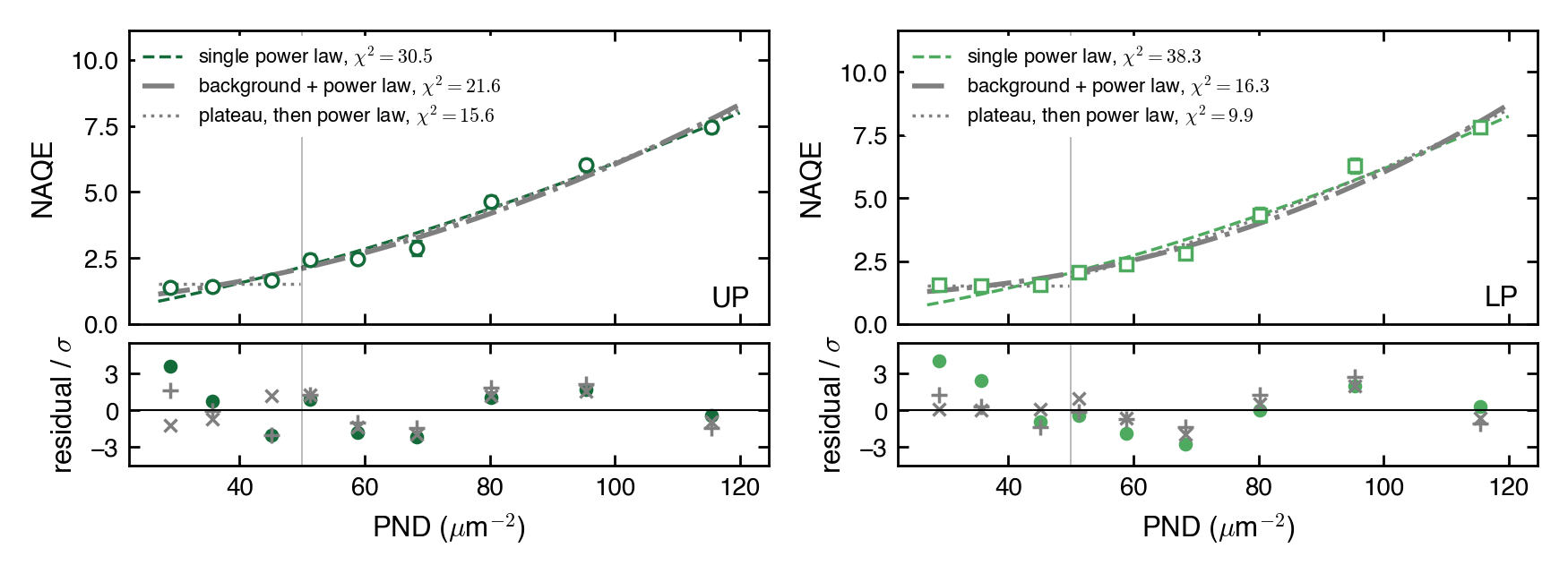}
	\caption{\textbf{Fit models for the polariton NAQE.} Fits of the NAQE of the upper (left) and lower (right) polariton with a single power law over all densities (dashed, as in Fig.~\ref{fig:AQEvsN}), a constant background plus a power law (dash-dotted), and the low-density plateau followed by a power law above ${\rm PND}=50~\mu$m$^{-2}$ (dotted). Lower panels show the residuals in units of the uncertainty.}
	\label{fig:fits}
\end{figure*}
The eigenvalues of the reconstructed matrix reproduce the four measured quantities exactly. This gives $G\propto{\rm PND}^{0.73\pm0.03}$, from 84 to 241~meV (Fig.~\ref{fig:G_hypothesis}(a)), and a cooperativity growing from $1.6$ to $15$ (Fig.~\ref{fig:spectral}). The cavity loss falls from 357 to 216~THz, as expected if the array forms the top mirror. The bright-mode width grows from 114 to 164~THz, $\gamma_B\propto{\rm PND}^{0.28\pm0.09}$, while the AT linewidth grows from 290 to 360~THz. The sum $\kappa+\gamma_B$ is fixed by the data, whereas its split depends on the width extraction, and two-Lorentzian fits give erratic values of $\gamma_B$. The inversion assumes a single bright mode and $\Omega_B$ equal to the AT peak.

\subsection{Local versus collective hot-electron injection}
\label{sec:local_vs_collective}
\noindent
We compare the measured NAQE of the ATA structures with Eq.~\eqref{eq:phi_drive}, which separates the optics of the structure from the transfer rate $N_{\rm coh}\Gamma$. Local transfer corresponds to $N_{\rm coh}=1$ and collective transfer to $N_{\rm coh}>1$.

\textit{Mirror absorption.} The two polaritons drain the same bright mode, so they share $N_{\rm coh}\Gamma$, $\gamma_B$ and the normalisation of the AQE. They differ only in the pump detuning $\delta_\pm=\omega_\pm-\Omega_B$, which is two to four times larger for the lower polariton. Since the NAQE is proportional to $\phi_{\rm ATA}$ with the same constant for both polaritons, the second factor of Eq.~\eqref{eq:phi_drive_factor} cancels in their ratio
\begin{equation}
	\frac{{\rm NAQE}_{\rm LP}}{{\rm NAQE}_{\rm UP}}=\frac{\kappa_0\gamma_B(1+4\delta_+^2/\gamma_B^2)+4G^2}{\kappa_0\gamma_B(1+4\delta_-^2/\gamma_B^2)+4G^2}
	\label{eq:LP_UP_ratio}
\end{equation}
which depends only on $\kappa_0$ and on the spectral parameters of Sec.~\ref{sec:fit_ata}, and equals one for $\kappa_0=0$. The measured ratio is $1.01\pm0.03$ on average, with $\chi^2=6.8$ for nine densities at $\kappa_0=0$. A mirror absorption of 7.6 (15)~THz would lower the LP yield by up to 20\% (30\%) and raise $\chi^2$ to 18 (44). The data bound $\kappa_0<4.4$~THz at $2\sigma$, the largest $\kappa_0$ with $\chi^2$ within 4 of its minimum, about 2\% of $\kappa$ (2.7~THz with Lorentzian widths). The first factor of Eq.~\eqref{eq:phi_drive_factor} is therefore close to one, and ${\rm NAQE}\cdot\gamma_B\propto N_{\rm coh}\Gamma$.
\textit{Transfer rate.} At each density $i$ we take $Y_i={\rm NAQE}_i\cdot\gamma_{B,i}$. We take the uncertainty of $\gamma_B$, which has no error bars, as its relative scatter $s=12\%$ around a power law in PND, and add it in quadrature to that of the NAQE,
\begin{equation}
	\sigma_i=Y_i\sqrt{(\sigma_{{\rm NAQE},i}/{\rm NAQE}_i)^2+s^2}.
\end{equation}
For a model $f$ with $n$ parameters fitted to the nine values $Y_i$,
\begin{equation}
	\chi^2=\sum_{i=1}^{9}\frac{[Y_i-f({\rm PND}_i)]^2}{\sigma_i^2},
\end{equation}
with $\nu=9-n$ degrees of freedom. The $p$-value is the probability that a $\chi^2$ variable with $\nu$ degrees of freedom exceeds the measured value. Fitting ${\rm NAQE}\cdot\gamma_B\propto{\rm PND}^{\alpha_\Gamma}$ over all nine densities gives $\alpha_\Gamma=1.68\pm0.13$ for the upper and $1.73\pm0.14$ for the lower polariton (Fig.~\ref{fig:G_hypothesis}(b)), with $\chi^2=20.6$ and $24.6$ for 7 degrees of freedom (nine points, two parameters). Local transfer, a constant with one parameter, the weighted mean of the $Y_i$, gives $\chi^2=134.1$ and $118.2$ for 8 degrees of freedom. The corresponding $p$-values are $4\times10^{-25}$ and $8\times10^{-22}$ for local transfer, against $4\times10^{-3}$ and $9\times10^{-4}$ for the power law. With a constant $\gamma_B$, $\alpha_\Gamma$ would equal the NAQE exponents, $1.49\pm0.05$ and $1.60\pm0.05$. Figure~\ref{fig:fits} compares the single power law of Fig.~\ref{fig:AQEvsN} with a constant background plus a power law, and with the low-density plateau followed by a power law above ${\rm PND}=50~\mu$m$^{-2}$.

\textit{Assumptions.} The analysis assumes (i)~a single bright mode with uniform rates, (ii)~$\Omega_B$ equal to the AT peak, (iii)~peak widths that approximate the eigenvalue widths, (iv)~an AQE proportional to the yield per photon dissipated, and (v)~a transfer rate per particle that does not depend on density. The value of $\alpha_\Gamma$ depends on the split of $\kappa+\gamma_B$, whereas $\alpha_\Gamma\ge\alpha$ holds whenever $\gamma_B$ does not decrease with density. A local mechanism could mimic the measured growth only with a rate per particle that increases several-fold with density in the cavity alone, while the AT yield stays flat.

\section{Measurement protocol and separation of transfer from optics}
\label{app:em}
\begin{figure}[b]
    \centering
    \includegraphics[width=0.95\linewidth]{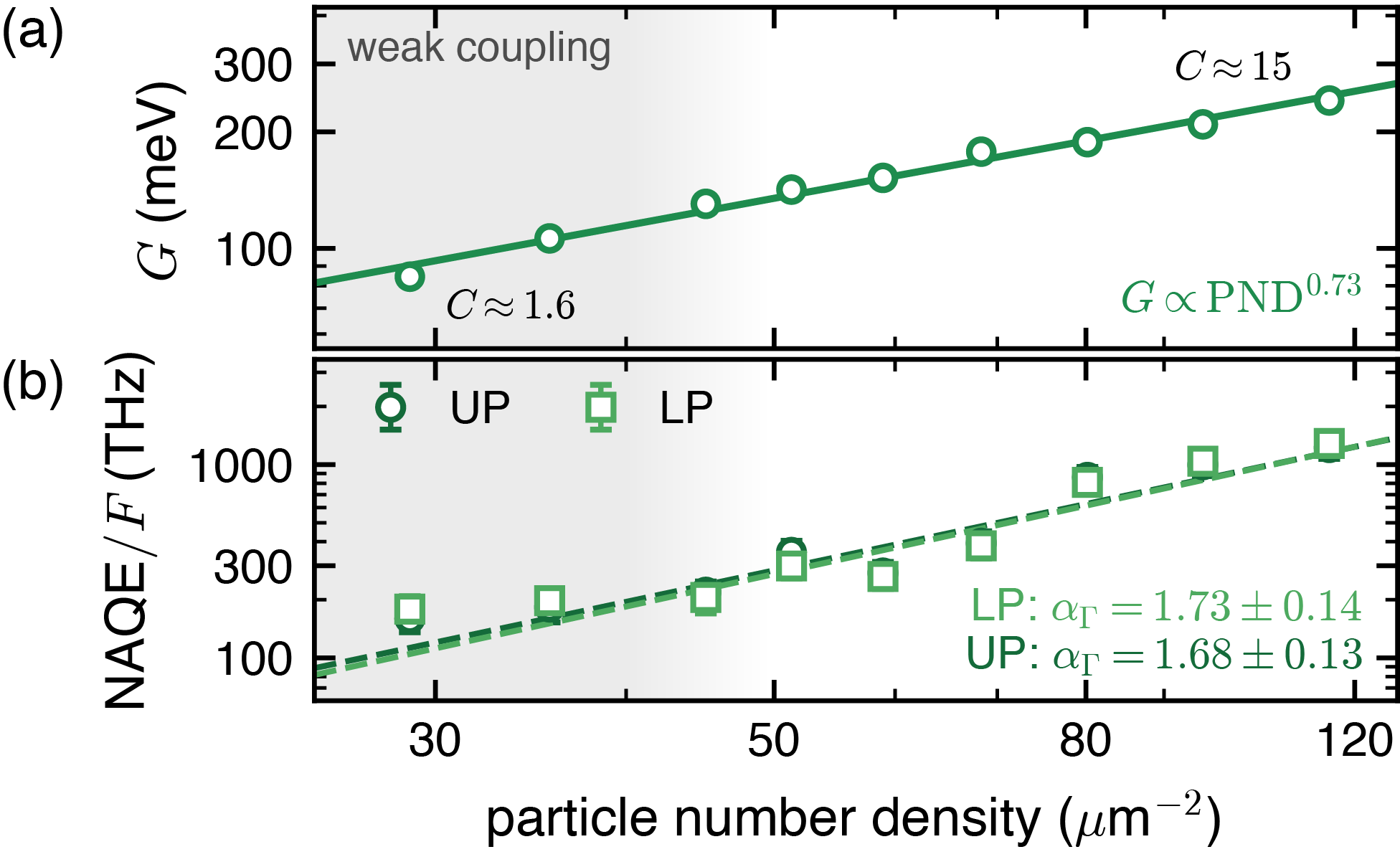}
	\caption{\textbf{Separating hot-electron transfer from optics.} (a)~Collective coupling $G$ from the ATA spectra (circles), with a power-law fit (solid line). The cooperativity $C=4G^2/\kappa\gamma_B$ grows from about $1.6$ to $15$. (b)~${\rm NAQE}\cdot\gamma_B$, proportional to the transfer rate $N_{\rm coh}\Gamma$ of the bright mode, for the upper (dark green circles) and lower (light green squares) polariton. Error bars include the uncertainty of $\gamma_B$. Dashed lines are power-law fits over all densities, as in Fig.~\ref{fig:AQEvsN}. Local transfer would give a constant. The shaded region marks weak coupling regime, where the metal nanoparticle density is so low that they hardly form the top mirror of the cavity.}
    \label{fig:G_hypothesis}
\end{figure}

\paragraph*{Measurement protocol.}
\label{em:protocol}
Photoexcitation was performed using resonant visible pump pulses tuned to the collective plasmon (AT case) or polariton resonance (both upper and lower branches, ATA case), while the injected electron population in TiO$_2$ was monitored by time-resolved mid-infrared transient absorption at a probe wavelength of 3~$\mu$m.
The peak transient signal, $\Delta\mathrm{OD}_{\max}$, is proportional to the conduction-band electron population.
The reported linear scaling with incident pump power confirms single-excitation regime operation~\cite{Liu_AN2023a}.

\paragraph*{Growth of the coupling with density.}
\label{em:G}
The data of Ref.~\cite{Liu_AN2023a} indicate that $G$ scales faster than $\sqrt{N}$, as shown in Fig.~\ref{fig:G_hypothesis}~(a). Here, we propose that this scaling reflects the fact that the metal nanoparticles form the top mirror of the cavity, so that the coupling $g$ of each particle grows with density before saturating. At low density, the particles are weakly coupled and barely form a mirror, leading to low delocalisation $N_{\rm coh} < N$. 

\paragraph*{Separating transfer from optics.}
\label{em:separating}
To test the two hypotheses of local and collective electron capture, we must separate the electron transfer contribution to the yield of Eq.~\eqref{eq:phi_drive}, given by $N_{\rm coh}\Gamma$, from the optical contribution. As discussed above, due to the low mirror absorption $\kappa_0$, we obtain $\phi_{\rm ATA} \gamma_B \approx N_{\rm coh} \Gamma$. By estimating $\gamma_B$ directly from the polaritons spectra, we isolate the transfer contribution of the normalised apparent quantum yield, shown in Fig.~\ref{fig:G_hypothesis}~(b), expressed as ${\rm NAQE}\cdot\gamma_B$; see Appendix~\ref{sec:local_vs_collective} for details. The latter must be constant if transfer is local, and is expected to grow with $N_{\rm coh}$ if transfer is collective. We fit ${\rm NAQE}\cdot\gamma_B\propto {\rm PND}^{\alpha_{\Gamma}}$ for both polaritons across all densities, obtaining $\alpha_\Gamma = 1.68\pm0.13$ for the UP and $\alpha_\Gamma = 1.73\pm0.14$ for the LP. Testing the two hypotheses against the UP (LP) data of Fig.~\ref{fig:G_hypothesis} we obtain $\chi^2/\nu = 16.8$ $(14.8)$ for the \textit{local} transfer hypothesis and $\chi^2/\nu = 2.9$ $(3.5)$ for the \textit{collective} transfer hypothesis, with $\nu$ being the number of degrees of freedom, equal to the difference between the number of data points and fit parameters. The corresponding $p$-values, the probability of obtaining a $\chi^2$ at least as large if the hypothesis were true, are $p \approx 10^{-25}$ $(10^{-21})$ for local transfer, which is therefore excluded, and $p \approx 5\times10^{-3}$ $(10^{-3})$ for collective transfer. Collective transfer thus describes the data far better, although not within the stated uncertainties. 

\end{document}